%% file: paper.tex
%
%
%

\ifx\mnmacrosloaded\undefined \input mn\fi

\input fig000
\figmod{1}                      


\pageoffset{-2.5pc}{0pc}

%
  
%
%
%

\Autonumber  


\pagerange{0-0}    
\pubyear{0000}
\volume{000}
%
%
\def\HI{\hbox{H~$\scriptstyle\rm I\ $}}

\def\lunits{\,{\rm W\,Hz^{-1}\,Mpc^{-3}}}
\def\iunits{\,{\rm erg\,cm^{-2}\,s^{-1}\,Hz^{-1}\,sr^{-1}}}

\def\Lya{Ly$\alpha\ $}
\def\etal{{et al.\ }}
\def\spose#1{\hbox to 0pt{#1\hss}}
\def\lta{\mathrel{\spose{\lower 3pt\hbox{$\mathchar"218$}}
     \raise 2.0pt\hbox{$\mathchar"13C$}}}
\def\gta{\mathrel{\spose{\lower 3pt\hbox{$\mathchar"218$}}
     \raise 2.0pt\hbox{$\mathchar"13E$}}}
\def\U3{U_{300}}
\def\B4{B_{450}}
\def\V6{V_{606}}
\def\I8{I_{814}}
\def\ub{U_{300}-B_{450}}
\def\uv{U_{300}-V_{606}}
\def\uve{(U_{300}-V_{606})_{\rm eff}}

\def\ui{U_{300}-I_{814}}
\def\bv{B_{450}-V_{606}}
\def\vi{V_{606}-I_{814}}
\def\bi{B_{450}-I_{814}}
\overfullrule=0pt

\begintopmatter  

\title{High Redshift Galaxies in the Hubble Deep Field: II. Colours and
Number Counts}
\author{Lucia Pozzetti$^{1,2,3}$, Piero Madau$^{3}$, Gianni Zamorani$^{2,4}$,
 Henry C. Ferguson$^{3}$, \& Gustavo A. Bruzual $^5$}
\smallskip
\affiliation{$^1$Dipartimento di Astronomia, Universit\`a di Bologna,
via Zamboni 33, I-40126 Bologna, Italy (lucia@astbo3.bo.astro.it)}
\smallskip
\affiliation{$^2$Osservatorio Astronomico di Bologna,
via Zamboni 33, I-40126 Bologna, Italy}
\smallskip
\affiliation{$^3$Space Telescope Science Institute, 3700 San Martin Drive,
Baltimore, MD 21218, USA}
\smallskip
\affiliation{$^4$Istituto di Radioastronomia del CNR, via Gobetti 101, I-40129 Bologna, Italy}
\smallskip
\affiliation{$^5$Centro de Investigationes de Astronom{\'\i}a, A.P. 264, 
M\'erida 5101-A, Venezuela}

\shortauthor{Pozzetti, Madau, Zamorani, Ferguson, \& Bruzual}
\shorttitle{High Redshift Galaxies in the HDF}


\acceptedline{Accepted 0000 xxxxxxxx 00. Received 0000 xxxxxxxx 00;
 in original form 0000 xxxxxxx 00}

\abstract {
We discuss the deep galaxy counts from the {\it Hubble Deep Field} (HDF)
imaging survey. At faint magnitudes, the slope of the differential number-magnitude
relation is flatter than $0.2$ in all four HDF bandpasses. In the ultraviolet,
a fluctuation analysis shows that the flattening observed below $U_{300}\approx 
26$ mag is
not due to incompleteness and is more pronounced than in the other bands,
consistent with the idea that a redshift limit has been reached in the galaxy
distribution. 
A reddening trend of $\approx 0.5$ magnitude is observed at faint
fluxes in the colour-magnitude diagram, $(\uv)_{eff}$ versus $V_{606}$. We
interpret these results as the effect of intergalactic attenuation on distant
galaxies. At flux levels of $AB\approx 27$ mag and in agreement with the
fluctuation analysis and the colour-magnitude relation, about $7$\% of the
sources in $U_{300}$, $30\%$ in $B_{450}$ and $35\%$ in $V_{606}$ are
Lyman-break ``dropouts'', i.e. candidate star-forming galaxies at $z>2$. By
integrating the number counts to the limits of the HDF survey we find that the
mean surface brightness of the extragalactic sky is dominated by galaxies that 
are relatively bright and are known to have $\langle z\rangle\sim0.6$. 
To $AB\approx 29$ mag, the integrated light from
resolved galaxies in the $I$-band is $2.1 ^{+0.4}_{-0.3} \times
10^{-20}\iunits$, and 
its spectrum is well described by a broken power-law ($I_\nu\propto\lambda^2$ 
from 2000 to 8000$\,$\AA\ and $I_\nu\propto \lambda$ from 8000 to 
$22000$ \AA). We discuss the predictions 
for the counts, colours, and luminosity densities
from standard low-$q_0$ pure-luminosity-evolution models
without dust obscuration, 
and find that they are unable to reproduce all the observed
properties of faint field galaxies. 
} 

\maketitle 

\section{Introduction}

As the best view to date of the optical sky at faint flux levels, the {\it
Hubble Deep Field} (HDF) imaging survey has rapidly become a key testing ground
for models of galaxy evolution. With its depth -- reaching 5-$\sigma$ limiting
AB magnitudes of roughly 27.7, 28.6, 29.0, and 28.4 in the F300W, F450W, F606W,
and F814W bandpasses (Williams \etal 1996) -- and four-filter strategy in order
to detect Lyman-break galaxies at various redshifts, the HDF offers the
opportunity to study the galaxy population in unprecedented detail. In
particular, the great cosmological importance of measuring the number-magnitude
relation (and its first moment, the integrated extragalactic background light)
and the colour distribution at the faintest detection limits reached by the HDF
stems from the prospect of using them as a probe of the cosmic history of star
formation in galaxies (Metcalfe \etal 1996; Ferguson \& Babul 1997). 

\fign\beginfigure*{\fignumber}
\putfigl{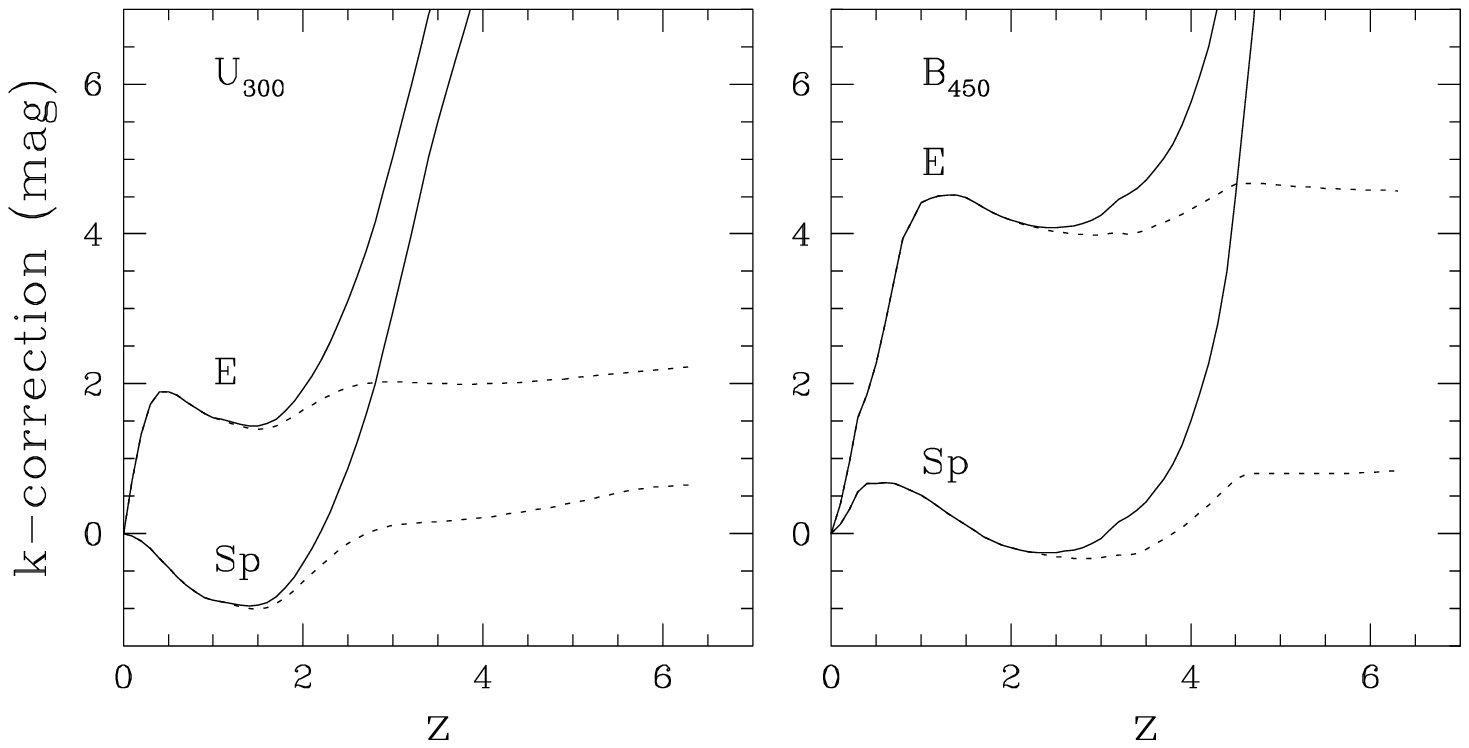}{7}{0}
\caption{{\bf Figure 1.}
Average $k$-corrections in the $U_{300}$ and $B_{450}$ bands for elliptical and
late-type spiral galaxies  as a function of redshift. {\it Dotted lines:}
unattenuated synthetic spectra. {\it Solid lines:} spectra modified by
intergalactic absorption.} 
\endfigure

At faint magnitudes, the interpretation and detailed modeling of the
observations require the self-consistent inclusion of the effect of
intergalactic attenuation on galaxy counts and colours.
Absorption by intervening material has been known for quite some time to
distort our view of objects at cosmological distances. It has been realized
only recently, however, that the increasing opacity of the intergalactic medium
(IGM) at high redshifts can be efficiently used to identify galaxies at
$z\gta2$. As shown by Madau (1995) (see also Yoshii \& Peterson 1994), the
accumulated Lyman-continuum absorption from the \Lya forest clouds and
Lyman-limit systems along the path is so severe that galaxies beyond $z\approx
3$ become effectively undetectable in the ultraviolet. Similarly, at $z\approx
3.5$, the blanketing by discrete absorption lines in the Lyman series is so
effective that galaxies appear about 1 mag fainter in the blue. Ground-based
observations have used colour techniques which are sensitive to the presence of
a Lyman-continuum break superposed to an otherwise flat UV spectrum to identify
or set limits on the number of high redshift galaxies (Guhathakurta, Tyson, \&
Majewski 1990; Steidel \& Hamilton 1992; Steidel \etal 1996a). Specific colour
selection criteria which exploit the ubiquitous effect of intergalactic
absorption and provide a robust separation between high-redshift and
low-redshift galaxies  have been developed for the HDF bandpasses by Madau 
\etal (1996). 

The magnitude and cosmological importance of intergalactic absorption on galaxy
number counts can be effectively illustrated by including its effect in the
standard $k$-correction term needed to translate the galaxy magnitude at Earth
into its rest-frame value. Assuming no intrinsic luminosity evolution,
the mean $k$-correction is
$$
k(z)=-2.5\log\left[(1+z){L(\nu_{\rm em})\over L(\nu_{\rm obs})}\langle
e^{-\tau}\rangle \right], \eqno(1) 
$$
where $L(\nu_{\rm em})$ is the specific power emitted by a source at redshift 
$z$,
$\nu_{\rm obs}=\nu_{\rm em}/(1+z)$, and $\langle e^{-\tau}\rangle$ is the cosmic
transmission averaged over all lines of sight.
Figure 1 shows the average $k$-correction in the $U_{300}$ and
$B_{450}$ bands as a function of redshift for synthetic spectra of galaxies
(based on Bruzual \& Charlot's 1993 libraries) which well reproduce the colours
of present-day ellipticals and late--type spirals. The effect at high redshifts is huge.
Due to intergalactic attenuation alone, the average $k$-correction in the F300W
bandpass increases by more than 4 magnitudes between $z\approx 2$ and
$z\approx 3.5$, giving origin to a ``UV dropout''. In the F450W band, the
increase is only noticeable above $z\approx3$, but becomes very large above
$z\approx 4$ (producing a ``blue dropout''). Qualitatively, the inclusion of 
intergalactic attenuation in any galaxy evolution model will cause a flattening,
more pronounced in the bluest bands, of the slope of the number-magnitude
relation and a reddening of the ultraviolet-optical colours.

In this paper we will focus on the $N(m)$ relation in the $U_{300}$, $B_{450}$,
$V_{606}$, and $I_{814}$ HDF bandpasses. A fluctuation analysis on the
$U_{300}$ image allows us to estimate the slope of the ultraviolet number 
counts below the nominal detection limit. These are found to flatten 
significantly below 26 mag, while a reddening trend of $\approx 0.5$ 
mag is observed in the colour-magnitude diagram, $\uve$ versus $V_{606}$. 
Both these results are consistent with the idea that, at flux
limits fainter than 27 mag, high-$z$ galaxies make a significant 
contribution to the counts in the redder bands, in agreement with an 
extension to faint magnitudes of the ``dropout" colour technique. We compute 
the extragalactic background light (EBL) from discrete sources, and find that 
it can be well described by a broken power-law, $I_\nu \propto \lambda^2$ 
between 2000 and 8000 \AA\ , and $I_\nu \propto \lambda$ between 8000 and 
22000 \AA.  Relatively bright galaxies at the limit of 
existing spectroscopic surveys are responsible for about 60\%
of the sky brightness. Finally, we examine a representative ($H_0=50$ km 
s$^{-1}$ Mpc$^{-1}$, $q_0=0.05$) pure-luminosity-evolution (hereafter PLE)
model without dust obscuration, and show that this is unable to reproduce all 
the observed properties of faint field galaxies. In particular, it 
overpredicts the fraction of $z\gta3$
objects and fails to account for the steep observed trend of the comoving
luminosity density in the interval $0<z<1$ at all wavelengths.

Throughout this paper, all magnitudes will be given in the AB system (Oke
1974). 

\fign\beginfigure*{\fignumber}
\putfigl{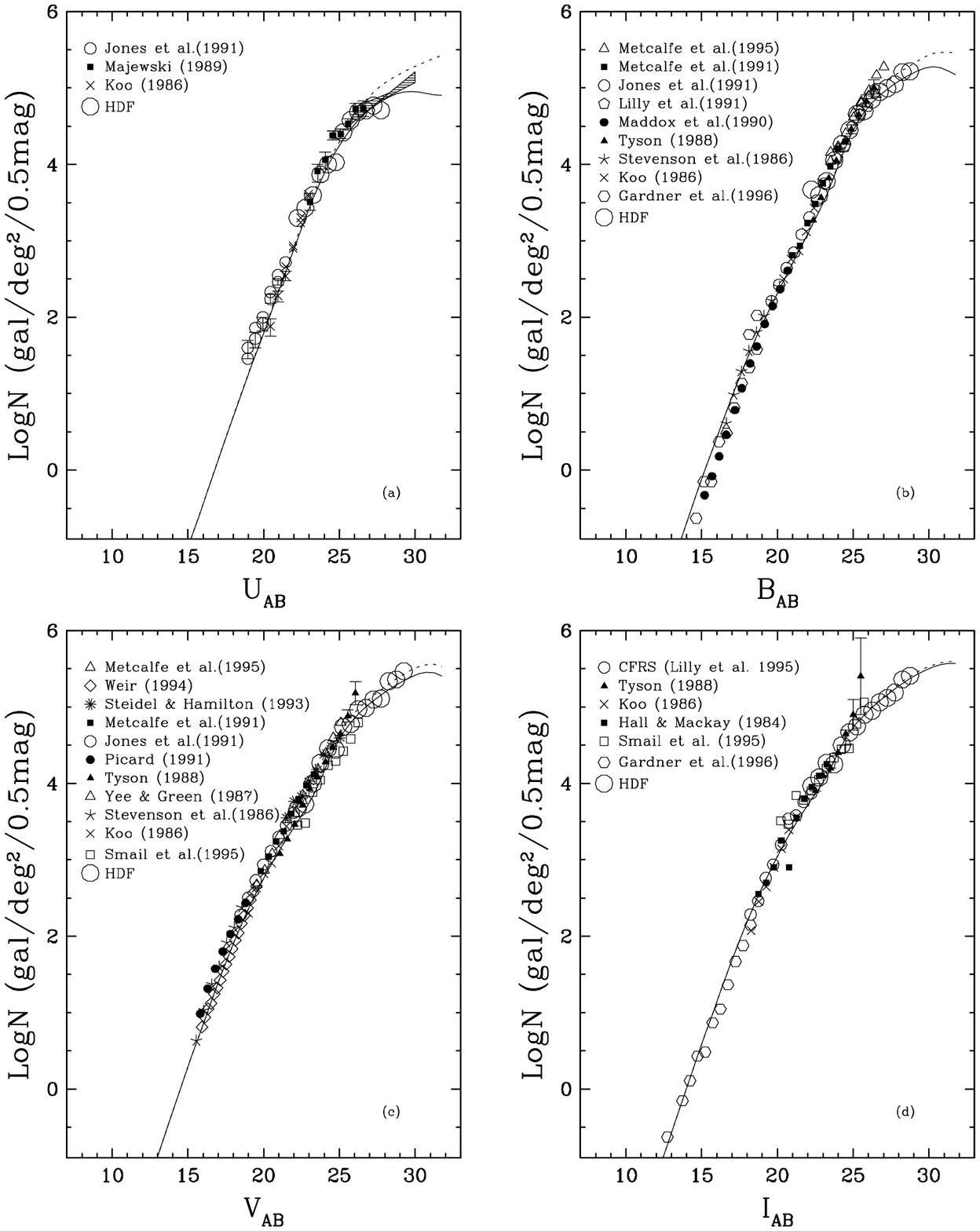}{20.9}{0}
\caption{{\bf Figure 2.}
Differential $U,B,V$, and $I$ galaxy counts as a function of AB magnitudes.
The sources of the data points are indicated in each panel. Note the decrease
of the logarithmic slope $d\log N/dm$ at faint magnitudes. The flattening is
more pronounced at the shortest wavelengths. The lines show the predicted
counts for a PLE model ($H_0=50$ Km s$^{-1}$ Mpc$^{-1}$, $q_0=0.05$), 
with ({\it solid line}) and without {\it (dotted line)} 
the inclusion of intergalactic absorption. The shaded region
in panel (a) shows the results of our ``fluctuation analysis" (see text).
}
\endfigure

\section {The Hubble Deep Field}

The HDF observations allow us to statistically analyze the high-$z$ galaxy 
population at magnitudes much fainter than current spectroscopic limits. In 
this section, we shall use the number counts, the Lyman-break dropout 
technique, and the colour-magnitude diagram to assess the effects of an 
absorption-induced loss of sources in the ultraviolet and blue bands. 

The galaxy sample we use, extracted from Version 2 of the HDF catalog, 
consists of $2819$ objects detected in the three WF chips from the F606W+F814W
summed images. The detection threshold was set to $4\sigma_{sky}$, 
and the minimum area to 25 drizzled pixels (Williams \etal 1996).
Magnitudes or upper limits in each band were computed from an area 
corresponding to the limiting isophote of the summed F606W+F814W image.
Details of the data reduction, source detection algorithm, and photometry are 
given in Williams \etal (1996), together with the angular separation and 
colour criteria adopted for the merging of subcomponents. 
It is important to keep in mind that the results discussed
below refer to a red-selected sample: in principle, objects with very blue 
colours, due for example to strong emission lines in the F300W or F450W 
bandpass, may have escaped detection if their red $V_{606}+I_{814}$ magnitudes
do not satisfy the selection criteria. 

Figure 2 shows the HDF galaxy counts as a function of AB isophotal magnitudes
in the F300W, F450W, F606W, and F814W bandpasses, as derived by
Williams \etal (1996) for all galaxies with signal-to-noise ratio $S/N>3$
within the band. At this limit the counts are likely to be more than $80\%$ 
complete.  A compilation of existing ground-based data is also shown.
All the surveys have been corrected to AB magnitudes, while the second order
colour corrections for the differences in the filter effective wavelengths 
have not been applied to the ground-based data. For the typical
colours of galaxies in the HDF, these corrections are less than 0.1 mag. 
The HDF counts agree reasonably well with previous surveys, to within $20\%$ 
in the magnitude range $22<AB<26$. At faint fluxes, however, the Williams et al.
counts in the four individual passbands are lower by about $30\%$ than those 
obtained for the HDF by Metcalfe et al. (1996).   
A detailed comparison of the Williams et al. and
Metcalfe et al. catalogs reveals several significant differences. First, there
are systematic discrepancies in the total magnitudes of the galaxies. At
$V_{606}= 25 \pm 0.5$, the median shift (Metcalfe$-$Williams) is $-0.01$ mag.
This grows to about $-0.20$ mag at fainter fluxes ($V_{606} = 28\div 29$),
and is probably due to different
algorithms used for ``growing" the photometry beyond the outer isophotes of the
galaxies. A further difference between the two catalogs is the amount of
splitting near bright objects, as these are split into slightly more pieces in
the Metcalfe et al. catalog than in Williams et al. Finally, the Metcalfe et
al. sample contains sources with lower S/N than the Williams et al.
sample. As the error on the magnitudes near the detection threshold is 
$\sim 0.5$ mag, the slightly fainter detection limit will tend to steepen 
the counts, even a magnitude or so above the nominal depth of the catalog. 
These three effects probably account for most of the discrepancies, 
and we believe are indicative of the possible systematic errors that are 
inherent in {\it HST} faint-galaxy photometry. 

It is important to note that, in all four HDF bands, the slope of the 
differential number-magnitude relation is flatter than 0.2 below $AB=26$ mag, 
as previously found by other authors (Williams \etal 1996; Metcalfe \etal 
1996), and that this flattening is more pronounced at the shorter wavelengths. 
We will show in Section 2.2, through a fluctuation analysis, that the turnover 
observed in the $U_{300}$ band is not due to incompleteness, but it is 
consistent with the idea that, at faint magnitudes in the red bands, a 
significant fraction of galaxies are actually at high redshifts, and are 
therefore ``reddened'' by absorbing material along the line of sight. 

\subsection{Biases}

We should caution the reader that there are a number of subtle effects
that can affect the photometry at very faint magnitudes, especially
in the $U_{300}$ band. First, the charge transfer efficiency (CTE) of the WFPC-2
CCD is not perfect, and can lead to magnitudes that are systematically too faint
for sources observed at low count rates with low background. 
For the typical background levels of 1 DN for the F300W exposures, the
average CTE correction is $\sim 0.03$ magnitudes for galaxies with 
$U_{300} \approx 25$. This correction will depend on the location of the source
on the chip, varying from zero at low row numbers to 0.5 mag near the top
of the chip.  The CTE effect on sources several magnitudes fainter
is not well calibrated, but is unlikely to be more than 0.1 mag. 
A second subtlety is that, at background levels of $\sim 1$ DN per exposure, the
determination of the local background for a galaxy is affected by the
discrete steps of the CCD analog-to-digital converter. While 
the algorithm used by FOCAS (a clipped mean) should be a good estimator of
the background, we have not tested this ourselves on simulated data.
Finally, for galaxy fluxes within several percent of the background, 
it is possible that systematic biases creep in due to the different treatment
of the pixel intensities in the source aperture and background region. 
In particular, the mean intensity in the local-background region is determined
after clipping out pixels more than 10 sigma away  from an initial estimate
of the mean, while the flux in the source aperture is an unclipped sum
of the counts. This could bias the fluxes of sources that are very faint in 
F300W towards slightly higher values, resulting in measured colours somewhat 
bluer
than the true colours. Finally, at magnitude levels fainter than $V_{606} = 28$,
it is likely that even objects which are brighter than the nominal depth of 
the catalog may be missed due to low surface brightness. These galaxies could be
preferentially either blue or red. For example, high redshift galaxies that are
not forming stars rapidly will be red and have quite low surface brightness.
The $(1+z)^4$ dimming effect is equal to 1.2 magnitudes between $z = 2$ and $z = 3$. On
the other hand, locally, LSB galaxies tend to be quite blue. A relatively local
population of LSB galaxies could in principle exist and the fainter examples
could disappear below the HDF sky background. 

With these caveats in mind, we will limit our analysis to galaxies with
$V_{606} < 29$ mag, with the expectation that systematic biases will be 
significantly less than $\sim 0.5$ mag in the photometry, and less than 
$\sim 50$\% in the galaxy counts. 

\subsection{Fluctuation Analysis}

As the F300W bandpass is sensitive to intergalactic attenuation at $z\gta2$,
deep counts in this band are essential for estimating, through the comparison 
with counts at longer wavelengths, the number of high-$z$ galaxies in a 
red-selected sample, and for deriving constraints on models of galaxy evolution.

\fign\beginfigure{\fignumber}
\putfigl{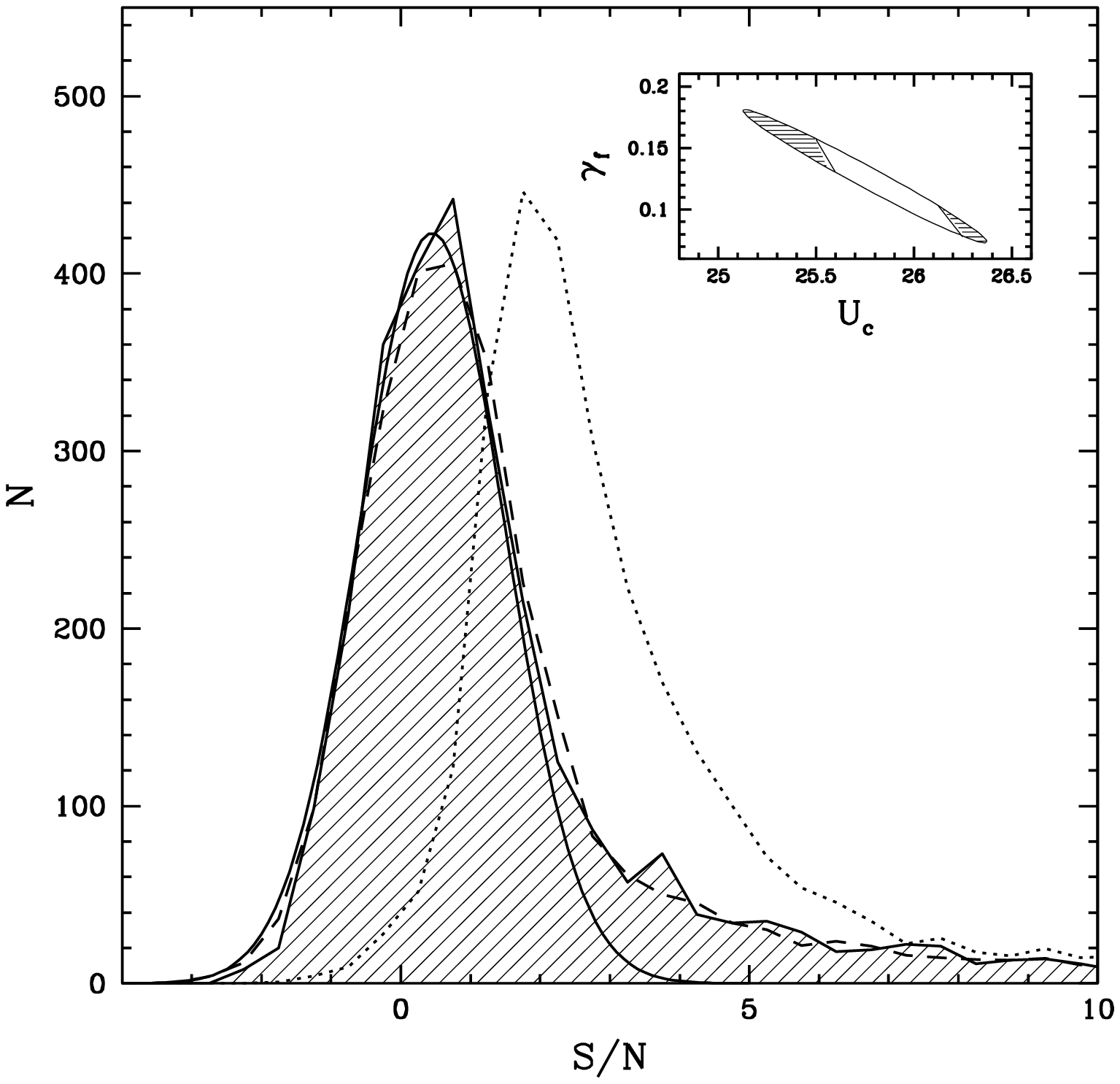}{9}{0}
\caption{{\bf Figure 3.}
S/N distribution in the $U_{300}$ band. The data ({\it shaded region}) are 
compared to a simulated distribution with $U_c=25.75$ mag, and 
$\gamma_f=0.135$ ({\it dashed line}) or 0.350 ({\it dotted line)}.
The gaussian curve represents the expected distribution for undetected objects
(see text).
The upper right panel shows the $95\%$ confidence region for the
parameters $U_c$ and $\gamma_f$ from a KS test (see text).
}
\endfigure

The published UV $N(m)$ (Williams \etal 1996) already suggests a sharp
flattening from a logarithmic slope of $d\log N/dm\approx 0.40$ at
$U_{300}\le26$ mag to a slope consistent with zero for $26<U_{300}<28$. In
principle, however, this sharp flattening could also be due to significant
incompleteness in the faintest $U_{300}$ bins. Note that,
since the limiting ultraviolet flux is about one magnitude brighter
than in the other bands, the fraction of galaxies in the catalog that are 
undetected (i.e. $S/N<3$) in the F300W bandpass is very large
($\sim 80\%$), and it is difficult to draw statistically reliable conclusions
from the raw data. 

In the attempt to push the $U_{300}$ counts to even fainter magnitudes, we have
therefore applied a ``fluctuation analysis'' to the Version 2 catalog data. The 
simplest possible model for the differential counts has been assumed, 
i.e. a double power-law
characterized by four parameters: the overall normalization ($A$), two slopes
($\gamma_b$ and $\gamma_f$ at bright and faint magnitudes, respectively), and
the magnitude ($U_c$) of the turnover. While two of them ($A$ and
$\gamma_b$) have been kept fixed by requiring a good fit to the bright HDF
$N(m)$, we have determined a best estimate for the other two parameters
($\gamma_f$ and $U_c$) in the following way: 

a) values for the ultraviolet magnitudes of the sources have been randomly 
extracted from the input $N(m)$;

b) the corresponding ``measured'' counts for each galaxy have been derived 
assuming Poisson statistics;

c) an isophotal area has been assigned to each object, on the basis of the
correlation between area and magnitude observed in the catalog, and the
corresponding number of background counts has been derived; 

d) finally, a signal-to-noise ratio $[S/N=L_i/\sigma(L_i)]$ has been computed 
according to the formula (Note that in equation 6 
in Williams \etal 1996 there was a typo in the second and third terms):
$$
\eqalign{[\Gamma \sigma(L_i)]^2= & \Gamma N_{obj} + 1.9^2 \Gamma^2
\sigma_{sky}^2 A_{obj} + \cr & 1.9^2 \Gamma^2 \sigma_{sky}^2 A_{obj}^2/A_{sky},}
\eqno(2) 
$$
where $L_i$ is the sky-subtracted number of counts within the detection 
isophote, $\Gamma$ is the inverse gain, $N_{obj}$ is the total number of
counts in the object aperture, $A_{obj}$ and $A_{sky}$ are the areas
within the source and sky apertures, and $\sigma_{sky}$ is the measured 
standard deviation of the background within the sky aperture 
(cf. Williams \etal 1996).

By repeating these steps a large number of times (10,000), we have constructed
a simulated distribution of signal-to-noise ratios for each input $N(m)$, and 
these distributions have then been compared to the observed one for all 
the $2819$ objects of the merged catalog. Figure 3 shows such a comparison 
for $U_c =25.75$. As expected, the observed distribution can be described as a 
gaussian at low $S/N$ values, which represents the expected distribution
for undetected objects,  plus a tail at high $S/N$ values.  Since these 
are computed not in random positions, but at the positions where galaxies are 
detected in the red bands, the gaussian--like distribution is not centered 
around zero, but is displaced toward a higher value ($S/N\sim 0.5$).
The two curves plotted in Figure 3 show the results of two simulations, 
corresponding to the slopes of $\gamma_f=0.380$ and $\gamma_f=0.135$. 
The case $\gamma_f\sim\gamma_b\sim0.4$ is clearly ruled out 
to a high degree of significance. From a KS statistical test the formal 
$95\%$ confidence region for $(U_c,\gamma_f)$ can actually be derived. 
The two parameters are obviously highly correlated, with steeper slopes 
allowed for brighter values of $U_c$ and vice-versa (see inset in the upper 
right corner of Figure 3).  Note, however, that the region in the upper left 
part of the 95\% confidence ellipse produces number counts which are lower 
than the observed ones for $25.75 \le
U_{300} \le 27.25$, while its lower right part produces counts which are higher
than the observed ones in the same magnitude range. By requiring the models 
to be consistent with the data to better than $5\%$ in the four bins between 
$25.75\le U_{300}\le 27.25$, the two hatched regions of the parameter space can
be further excluded, and the allowed ranges become $25.6<U_c<26.2$ and
$0.08<\gamma_f<0.16$.  The best-fit values are $U_c = 25.75$, and 
$\gamma_f = 0.135$. The  latter is steeper than the value of 0.05 quoted in 
Williams \etal (1996), but still flatter than the corresponding values for
the $B_{450}$, $V_{606}$, $I_{814}$ bands, as expected in the case of a 
significant absorption-induced loss of faint, distant 
sources. 

All the allowed models predict a number of galaxies between 20 to 30\% higher
than observed in the faintest bin quoted by Williams et al. (1996),
$27.5<U_{300}<28.0$. The Version 2 catalog contains $112$ galaxies in
this magnitude bin, with only $79$ satisfying the $S/N \ge 3$ constraint
used by Williams \etal in deriving their counts. This suggests some
incompleteness in the Williams \etal counts at this magnitude level. 
Since the catalog has been derived using isophotal magnitudes 
and the isophotal areas are defined in the red bandpasses, for galaxies with 
a given ultraviolet flux (i.e., with a given signal) there is an inverse 
correlation between $S/N$ and
isophotal area: smaller signal-to-noise ratios correspond to larger areas and 
therefore, in first approximation, to galaxies that are brighter in the red 
bands. For this reason, the sources eliminated because of the $S/N$ constraint 
are not a random sub-sample of the galaxies in this magnitude range, but 
correspond to those with the reddest $\uv$ and/or $\ui$ colours. 
This has to be taken into account if one analyzes the colour distribution of 
the galaxies in the faintest $\U3$ magnitude bin of the catalog.

%
\begintable{1}
\nofloat
\caption{{\bf Table 1.} Number of Lyman-break dropouts}
\halign{%
\hfil#\hfil & ~~\hfil#\hfil & ~~\hfil#\hfil \cr
\noalign{\vskip 3pt}\noalign{\hrule}\cr\noalign{\vskip 3pt}
~$V_{606}$ bin~ & ~Udrops~ & ~~~Bdrops~ \cr
\noalign{\vskip 3pt}\noalign{\hrule}\cr\noalign{\vskip 3pt}
23.75 & $~~5.0\pm1.9~$ & 0.0 \cr
24.25 & $~~6.0\pm2.4~$ & 0.0 \cr
24.75 & $~11.0\pm2.7~$ & 0.0 \cr
25.25 & $~16.9\pm3.3~$ & 0.0 \cr
25.75 & $~29.9\pm4.2~$ & $~1.0\pm1.0$ \cr
26.25 & $~35.4\pm5.3~$ & $~3.0\pm1.7$ \cr
26.75 & $~47.1\pm7.1~$ & $~2.5\pm1.6$ \cr
27.25 & $~68.5\pm9.2~$ & $~5.0\pm2.2$ \cr
27.75 & $ - $          & $12.8\pm6.8$ \cr
\noalign{\vskip 3pt}\noalign{\hrule}\cr\noalign{\vskip 3pt}
$23.5 \div 28.0$ & $219.8\pm14.6$  & $24.3\pm7.6$ \cr
\noalign{\vskip 3pt}\noalign{\hrule}\cr
}
\endtable

\fign\beginfigure{\fignumber}
\putfigl{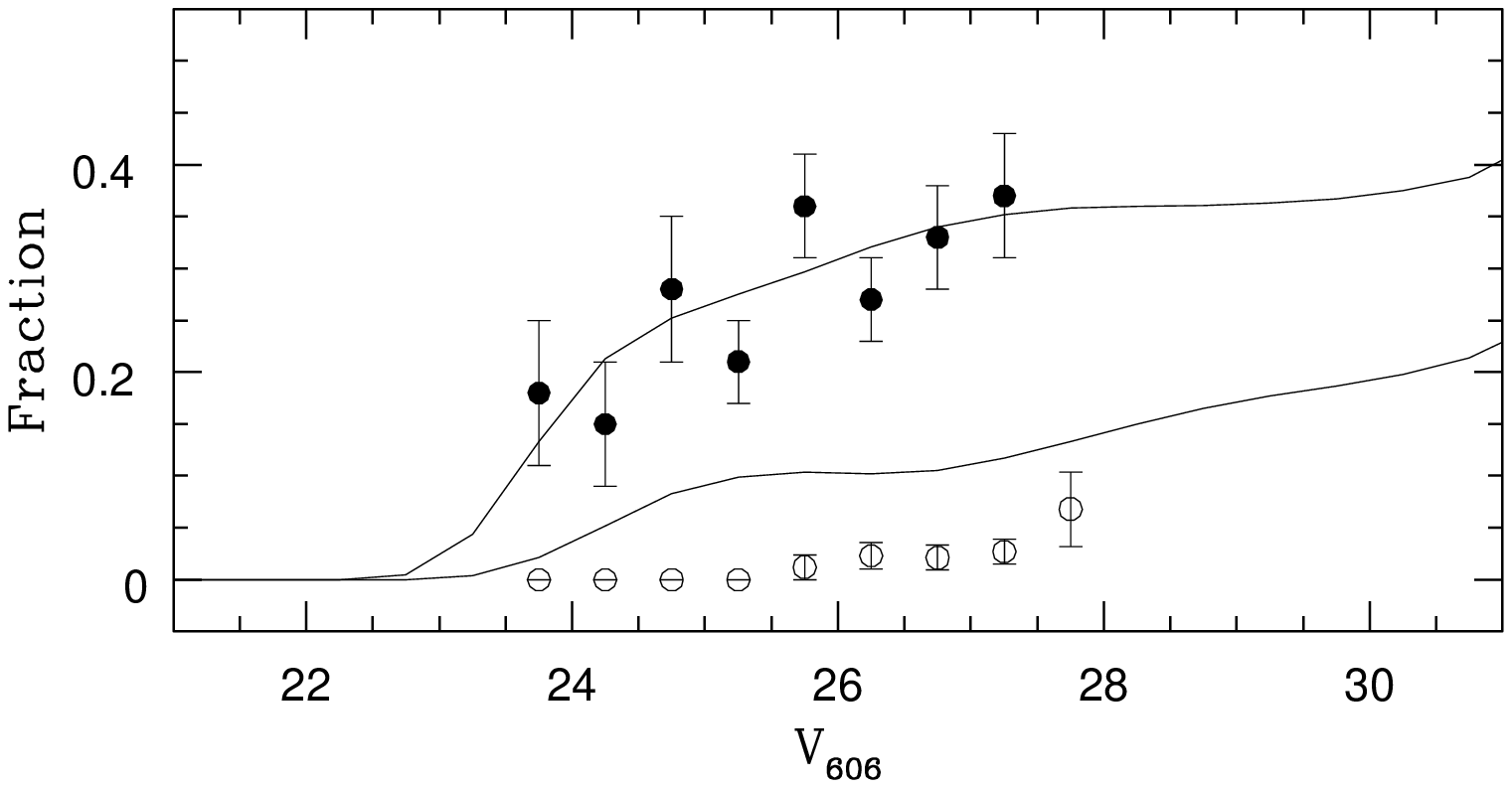}{5}{0}
\caption{{\bf Figure 4.} Fraction of high-$z$ galaxies as a function of
$V_{606}$ magnitudes as obtained from a dropouts analysis, using the
ASURV package. {\it Filled circles:} $U_{300}$-dropouts. {\it Empty
circles:} $B_{450}$-dropouts. {\it Solid lines:} PLE model predictions. 
}
\endfigure

As  mentioned  in Section 2, galaxies with very blue colours, not selected
in the catalog because fainter than the limits in $V_{606}+I_{814}$, could in 
principle contribute somewhat to the $\U3$ counts. Because our analysis uses as 
input the red selected catalog, this contribution would not be accounted for 
in our extrapolated estimate of the $\U3$ counts. Given the approximate limit 
in the $\V6$ band ($V_{lim} \sim 29.0$), these ``potential" blue objects could 
appear at $\U3 \sim 29$ if they had $\uv \le 0$ and at $\U3 \sim 29.5$ if they 
had $\uv \le 0.5$. In the ``bright" part of the catalog (i.e. $\V6 < 26$), 
where no such object is expected to have been lost, there is only $1\%$ of the 
galaxies with $\uv \le 0$ and only $12\%$ of the galaxies with $\uv \le 0.5$. 
On this basis we argue that our conclusion on the reality of the flattening in 
the $\U3$ counts is robust and our estimate of the parameters of the slope 
down to $\U3 \sim 29.5$ can be considered to be reliable.

\subsection{Lyman-Break Dropouts and the High-$z$ Luminosity Functions}

Photometric criteria for robustly selecting Lyman-break
galaxies in the HDF have been developed by Madau \etal (1996). They have been
tuned up to provide what appear to be largely uncontaminated samples of
star-forming galaxies at high redshifts, and have been refined (Madau 1997)
after the many redshift measurements with the Keck telescope (Steidel \etal
1996b; Cohen \etal 1996; Lowenthal \etal 1997). The HDF ultraviolet passband
-- which is bluer than the standard ground-based $U$ filter -- permits the 
identification of star-forming galaxies in the interval $2 \lta z \lta 3.5$. 
Galaxies in 
this redshift range predominantly occupy the top left portion of the $\ub$ vs.
$\bi$ colour-colour diagram because of the attenuation by the intergalactic medium
and intrinsic absorption. Galaxies at lower redshift can have similar $\ub$
colours, but are typically either old or dusty, and are therefore red in $\bi$ 
as well. In analogous way, it is possible to single out star-forming galaxies
at $3.5\lta z \lta 4.5$ by looking at the $\bv$ vs. $\vi$ colour-colour diagram.

We have estimated the number of $U_{300}$ and $B_{450}$ dropouts as a
function of $V_{606}$ magnitudes by applying, with some modifications, 
the procedure described in Madau \etal (1996). The main differences
with respect to Madau's procedure are: 
i) We considered as detected only objects with $S/N \ge 2.5$, rather than
$S/N \ge 1$ as in Madau's analysis. For the objects with a lower $S/N$ we
computed a lower limit to the magnitude  corresponding to $S/N=2.5$. As seen
in Figure 3, it is only for $S/N \ge 2.5$ that the observed distribution
starts to depart significantly from the expected distribution of the noise,
represented in the figure by the gaussian curve. Therefore, the objects
with lower $S/N$ can not be considered detections.
ii) Since with this choice we have a higher number of lower limits in the
colours, the number of dropouts in each $V_{606}$ bin has been estimated
using the Kaplan-Meier estimator in the package ASURV (La Valley,
Isobe \& Feigelson 1992), which implements the methods for a maximum-likelihood
reconstruction of the true distribution function when limits are present 
(Feigelson \& Nelson 1985). 
Table 1 shows the numbers of $U_{300}$ dropouts
for $V_{606}\le 27.5$ and of $B_{450}$ dropouts for $V_{606}\le 28$ as a
function of $V_{606}$ magnitudes. At fainter $V_{606}$ magnitudes the number
of limits in $U_{300}$ becomes greater than 60\% 
of the total number of data points. For these high percentages of limits the
performances of survival analysis statistics deteriorate and the results are
not reliable anymore (Feigelson 1992).
The corresponding fractions of dropouts are 
shown in Figure 4 as a function of $V_{606}$ magnitudes in $0.5$ magnitude bins.
The fraction of $U+B$ dropouts varies from $\sim 5\%$ to $40\%$ in the range  
$23.5<V_{606}<27.5$.
The cumulative fraction of F300W-dropouts to $V_{606}<27.5$ is $28\pm 2\%$, 
while it is $2.5\pm0.6\%$ to $\V6<28$ for the F450W-dropouts.

With a similar procedure we have computed the fraction of Lyman-break 
dropouts as a function of $U_{300}$ and $B_{450}$ magnitudes, finding that 
this decreases significantly in the $U_{300}$ band. For example, at $AB 
\approx 27$, it is $\sim 7\%$ in $U_{300}$, while it is  $\sim 30\%$ in 
$B_{450}$ and $\sim 35\%$ in $\V6$.

The Kaplan-Meier estimator used in this analysis can be properly applied only 
in presence of a random censorship, i.e. in cases in which the distribution
of the observational sensitivities for the non-detected objects is the same
as that for the detections (Schmitt 1985, Magri \etal 1988). In our case this 
condition is satisfied, because for each magnitude bin in $\V6$ the 
distribution of the thresholds in $U_{300}$ is essentially a function of 
$A_{obj}$ (see Eq. 2), which is not related to the variable to be measured 
(i.e. the $\U3$ magnitude).

However, the non-parametric Kaplan-Meier reconstruction of the true
distribution implicitely assumes also that the limits and the detections
are drawn from the same parent distribution. Since
we can not be sure that this assumption is true, we have analyzed the same
data also with a maximum likelihood parametric method. Consistently
with the analysis described in more detail in Section 2.4,
we have assumed the following functional form for the distribution of the
$U_{300} - B_{450}$ colours: a fraction $f$ of the population
of galaxies is highly absorbed in $\U3$ and is described formally by a 
$\delta$-function at a high value of $\ub$; a fraction $(1-f)$ of the
population is described by a gaussian distribution in $\ub$. The likelihood
for this functional form can be easily derived (see Avni \& Tananbaum 1986).
Maximizing the likelihood function one can then obtain
the best estimates for the three parameters ($f$ plus the two describing the
gaussian) and from these derive the best estimate for the number of dropouts,
i.e. those satisfying the condition of having a $\ub$ color greater than the
adopted threshold. The numbers derived with this procedure are all
consistent, within one sigma, with those given in Table 1 in each magnitude
bin. Since the assumptions in the two methods are significantly different,
this suggests that our estimate of the number of dropouts is statistically
robust.

\fign\beginfigure{\fignumber}
\putfigl{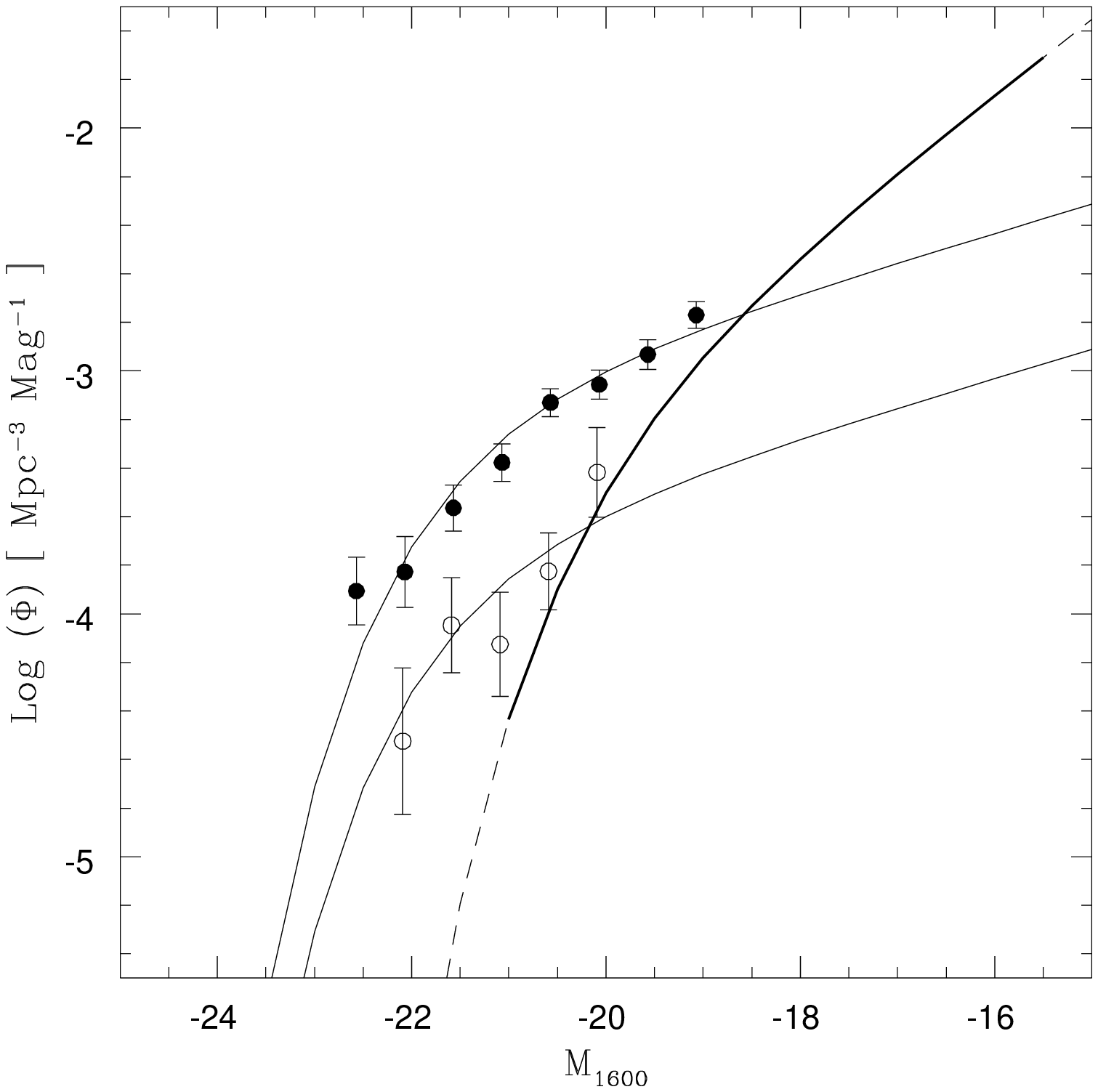}{8}{0}
\caption{{\bf Figure 5.} Luminosity function at $1600$ \AA\ ($H_0=50$ km 
s$^{-1}$ Mpc$^{-1}$, $q_0=0.05$). {\it Filled circles:} $U_{300}$-dropouts. 
{\it Empty circles:} $B_{450}$-dropouts. The solid lines represent Schechter
fits to the data. The local luminosity function at 2000 \AA\ 
from Treyer \etal (1997) is also shown ({\it thick curve}). 
}
\endfigure

As a test of the adopted colour selection window, we have compared our 
estimates of the fraction of bright UV dropouts in the HDF with the results
from the $B$-selected spectroscopic sample described by Cowie \etal (1997). 
The latter
covers, in the magnitude range $24<B<24.5$, about 13 arcmin$^2$, and contains
42 galaxies, of which 4 have $z\gta2$, and 2 are still unidentified (and most
likely lies at $z>1.6$, Cowie \etal 1997). The fraction of spectroscopically
identified galaxies at $z\gta2$ in the ground-based sample is therefore between
10 and 14\%. This is in good agreement with our estimated fraction, 
$\sim 11$\%, of ultraviolet dropouts in the same magnitude range. 

Using the numbers of dropouts as a function of magnitude, we have computed
the luminosity functions (LF) of high-$z$ galaxies in two different 
redshift ranges.
The observed $V_{606}$ and $I_{814}$ magnitudes have been converted to 
intrinsic $M_{1600}$ AB magnitudes, and the  number densities of $U_{300}$ and
$B_{450}$ dropouts have been obtained assuming that they sample the redshift 
intervals $2.0<z<3.5$ and $3.5<z<4.5$, respectively. The results are shown 
in Figure 5 ($H_0=50$ km s$^{-1}$ Mpc$^{-1}$, $q_0=0.05$). 
The shape of the LF for the redshift interval 
$2.0<z<3.5$ is reasonably well defined and
can be described by a Schechter function with parameters $\alpha=-1.3$,
$M_{1600}^*=-21.7$ and $\Phi^*=0.8\times 10^{-3}$ Mpc$^{-3}$.
For the higher redshift interval the fitting line shown in the figure has
the same $\alpha$ and $M_{1600}^*$ parameters, but a lower normalization 
($\Phi^*=0.2\times 10^{-3}$ Mpc$^{-3}$). As a comparison, the thick curve in 
Figure 5 shows the local $M_{1600}$ LF, derived from the 
Treyer \etal (1997) data. The high-$z$ luminosity
functions clearly extend to much brighter magnitudes, while no indication
of a steep faint end slope is clearly seen in the HDF data. Even the HDF 
data, however, are not deep enough to exclude the possibility of a steepening
at faint fluxes.
The comoving luminosity density resulting from the integration to $M_{1600} = 
-18$ of the two HDF LF are 
$\rho_L=2.0(\pm 0.5) \times 10^{19} \lunits$ 
for $2.0<z<3.5$, and $\rho_L=5.2(\pm 2.6) \times 10^{18} \lunits$ 
for $3.5<z<4.5$, in good agreement with the estimates of Madau (1997).
The errors on $\rho_L$ have been obtained by combining the
statistical errors on the number of dropouts (see Table 1) with an estimated
uncertainty of the volumes effectively sampled by the adopted colour
selection techniques.

These luminosity functions (and the corresponding luminosity densities)
should probably be considered as lower limits to the true ones. In fact, 
on the basis of an analysis of the spectra and colours of a representative, 
spectroscopically confirmed sample of high redshift star-forming galaxies, 
Pettini et al. (1997) have shown that the likely dust correction to the 
integrated ultraviolet luminosity for these galaxies is about a factor of 
3. (Note that the Treyer et al. LF has not been 
corrected for extinction). Moreover, the dropout technique used in selecting 
our high redshift candidates works
efficiently for galaxies for which the attenuation due to intergalactic
matter is equal to or higher than the assumed average attenuation. Galaxies with
a lower than average attenuation may well have $(U_{300} - B_{450})$ and 
$(B_{450} - V_{606})$
colours bluer than those used in selecting the high-$z$ sample. Madau \etal
(1996) have estimated that only about $20\%$ of galaxies at $z>2.7$  
can be missed because of this effect. 
However, only extensive spectroscopic observations
of galaxies lying below the colour threshold can provide a quantitative
estimate of the size of this incompleteness. 

\fign\beginfigure{\fignumber}
\putfigl{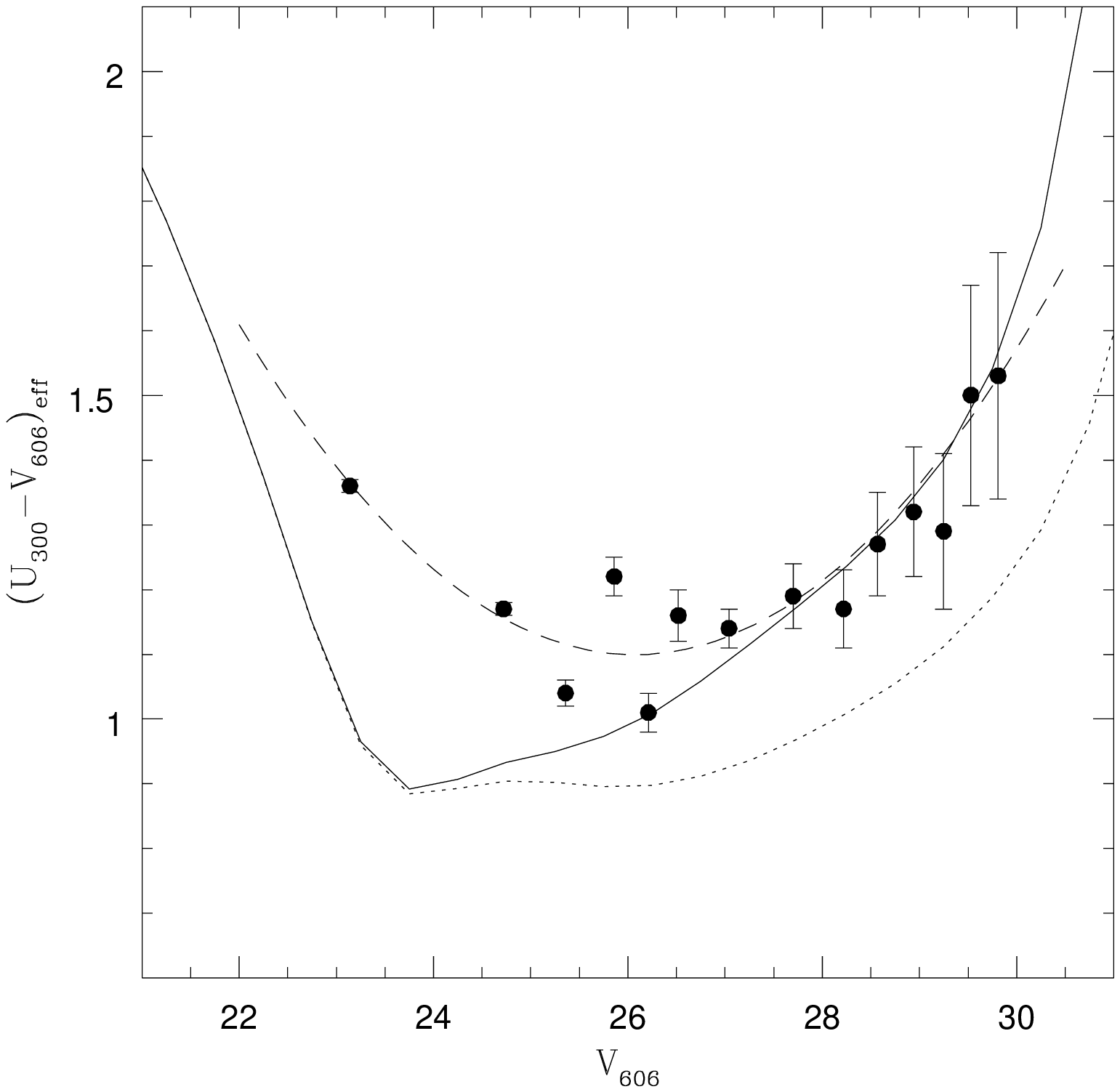}{8}{0}
\caption{{\bf Figure 6.} Colour-magnitude diagram, $(U-V)_{eff}$ vs. 
$V_{606}$ ({\it filled circles}).
The six brighter bins contain 80 galaxies each, while the fainter
bins contains 240 objects. The dashed curve represents a quadratic fit to the 
data to $V_{606}<29$. {\it Solid line}: prediction from a PLE model 
including the attenuation by the IGM. {\it Dotted line}: unattenuated PLE model.
}
\endfigure

\subsection{Colours}

As already mentioned, the number of objects detected in $U_{300}$ is 
a relatively small fraction of the total number of galaxies in the catalog. 
This obviously prevents a detailed analysis of the full $\uv$ distribution at
faint $V_{606}$ magnitudes, where for most of the objects only an upper limit
to the ultraviolet flux is available. 

It is possible, however, to use the measured fluxes after background 
subtraction for all the galaxies to get a statistical estimate on the average 
properties of the $\uv$ vs. $V_{606}$ relation at much fainter $V_{606}$ 
magnitudes than allowed by the $U_{300}$ detections. We have 
therefore computed in each
bin of $V_{606}$ magnitude the total F300W and F606W fluxes as the sum of all
the counts measured in individual galaxies, and have derived an
``effective'' colour for the entire population, $\uve$. This is
not equal to the mean value, $\langle \uv\rangle$. For example, if the colour
distribution were a gaussian with mean $\langle \uv\rangle$ and dispersion
$\sigma$, it is easy to derive the relation $\langle \uv\rangle=
\uve+0.4605\sigma^2$. 

\fign\beginfigure{\fignumber}
\putfigl{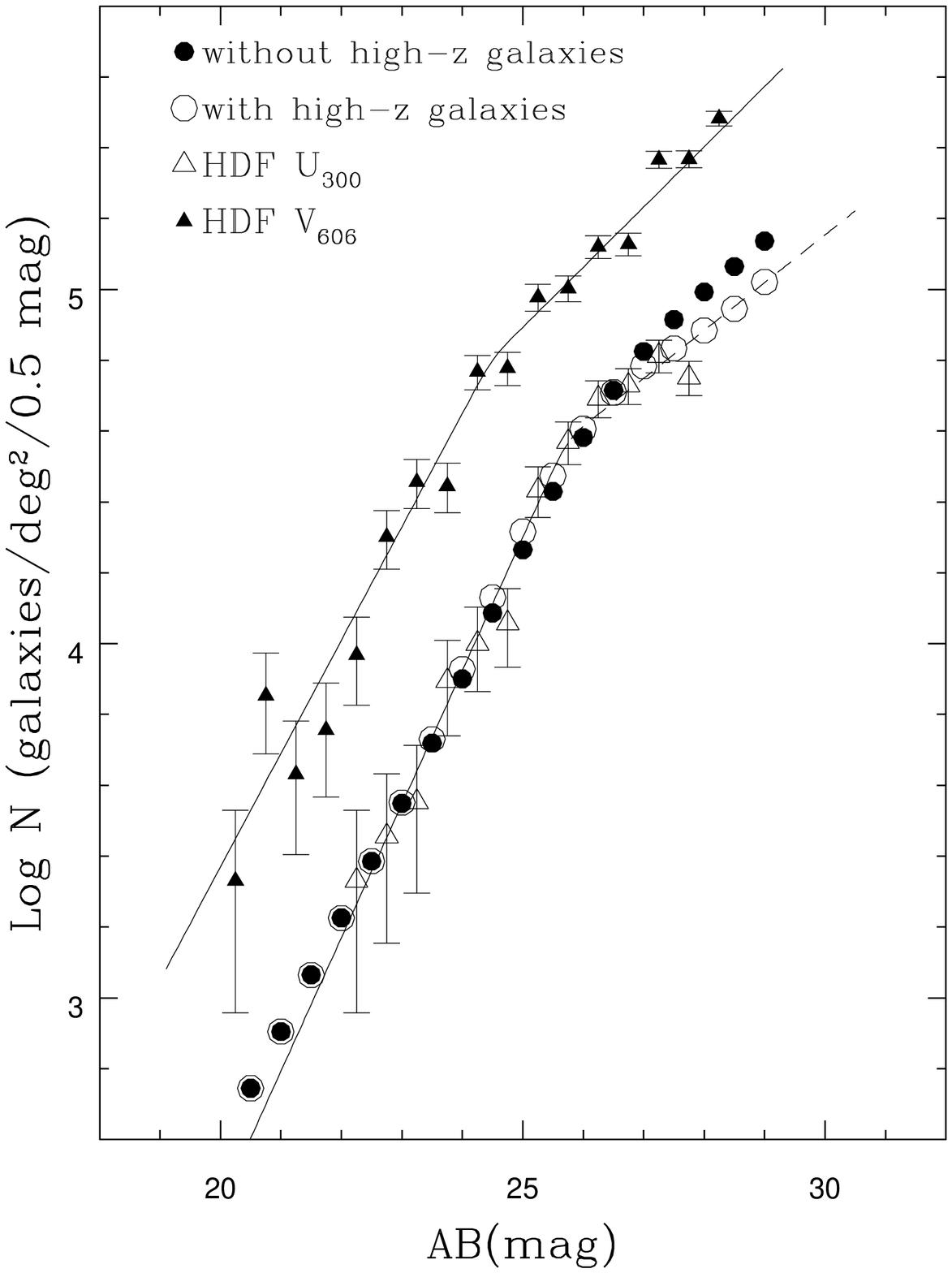}{10}{0}
\caption{{\bf Figure 7.} $U_{300}$ ({\it open triangles}) and $V_{606}$ 
({\it filled triangles}) number-magnitude relation from the HDF data. 
Simulated ultraviolet counts are also plotted, with ({\it open circles)} and 
without ({\it filled circles}) the inclusion of high-$z$ galaxies (see text). 
The solid lines depict the best-fits to the $V_{606}$ data and the 
$U_{300}$ $N(m)$ relation, with the former brightened by $1$ magnitude for 
clarity.}
\endfigure

Figure 6 shows the resulting colour-magnitude diagram. After an initial 
blueing trend, which is expected in PLE models (Pozzetti, Bruzual \& 
Zamorani 1996), at fainter magnitudes galaxies become redder in $\uve$. 
This may be due to an increase in the fraction of high redshift galaxies 
or to a change in the colour properties of low redshift objects. As a test 
for distinguishing between these two alternative possibilities we have 
estimated the expected ultraviolet counts from the observed optical counts
and the derived $\uve$ colours. The following input ingredients have been used: 

i) the observed $\log N(m)-V_{606}$ relation, which we fitted as a double
power-law with $\gamma_b=0.32$, $\gamma_f=0.17$, and $V_c=25.5$;

ii) an assumed gaussian distribution for $\uv$ with a mean such
that the resulting $\uve$ is equal to the fitted value in each $\V6$ bin;

iii) two different assumptions for the fraction, $f(V_{606})$, 
of high redshift galaxies in $0.5$ magnitude bins: 
$f_1(V_{606})=0$ and 
$f_2(V_{606})=0.097\times (V_{606}-22.68)$ for
$23 \le V_{606}\le27$, as obtained by the dropout technique (see \S~2.2).
At fainter flux levels, where the dropout analysis fails,
we have assumed that $f_2(V_{606})$ remains constant.

Figure 7 shows the results for the two cases. In the very simple approach 
followed here we have assigned the same colour, $\uv=4\,$mag, to all 
high redshift objects, which therefore contribute only a negligible fraction 
of the total measured $U_{300}$ flux in each $V_{606}$ bin. 
The $\uve$ colour for the
low redshift galaxies is then recomputed by scaling the total measured 
$V_{606}$ flux in each bin by $1-f_2(V_{606})$. In the same figure we show
the data points with error bars, the fit to the bright counts, 
and the fit at fainter magnitudes derived from the fluctuation analysis. 
The predicted ultraviolet $N(m)$ is significantly higher than
what is allowed by the fluctuation analysis for $f=f_1$, 
while it is in excellent agreement with the fluctuation analysis for 
$f=f_2(V_{606})$.

%
\begintable{2}
\nofloat
\caption{{\bf Table 2.} Integrated Galaxy Light}
\halign{%
\hfil#\hfil & \hfil#\hfil & \hfil#\hfil & ~~\hfil#\hfil & ~~\hfil#\hfil \cr
\noalign{\vskip 3pt}\noalign{\hrule}\cr\noalign{\vskip 3pt}
~~$\lambda$ (\AA) & ~~$AB (range)$~~ & ~~$I_{\nu}^a$~~ & 
~~$\sigma^+$~~ & ~~$\sigma^-$~~ \cr
\noalign{\vskip 3pt}\noalign{\hrule}\cr\noalign{\vskip 3pt}
~$3600$     & $18.0$--$28.0$ & $0.35$ & 0.07 & 0.05 \cr
~$4500$     & $15.0$--$29.0$ & $0.56$ & 0.11 & 0.07 \cr
~$6700$     & $15.0$--$29.5$ & $1.49$ & 0.27 & 0.20 \cr
~$8100$     & $15.0$--$29.0$ & $2.10$ & 0.44 & 0.25 \cr
$22000$     & $12.0$--$25.5$ & $5.81$ & 1.50 & 0.89 \cr
\noalign{\vskip 3pt}\noalign{\hrule}\cr
}
\tabletext{\noindent
$^a$ in units of $10^{-20} \iunits$.}
\endtable

\fign\beginfigure{\fignumber}
\putfigl{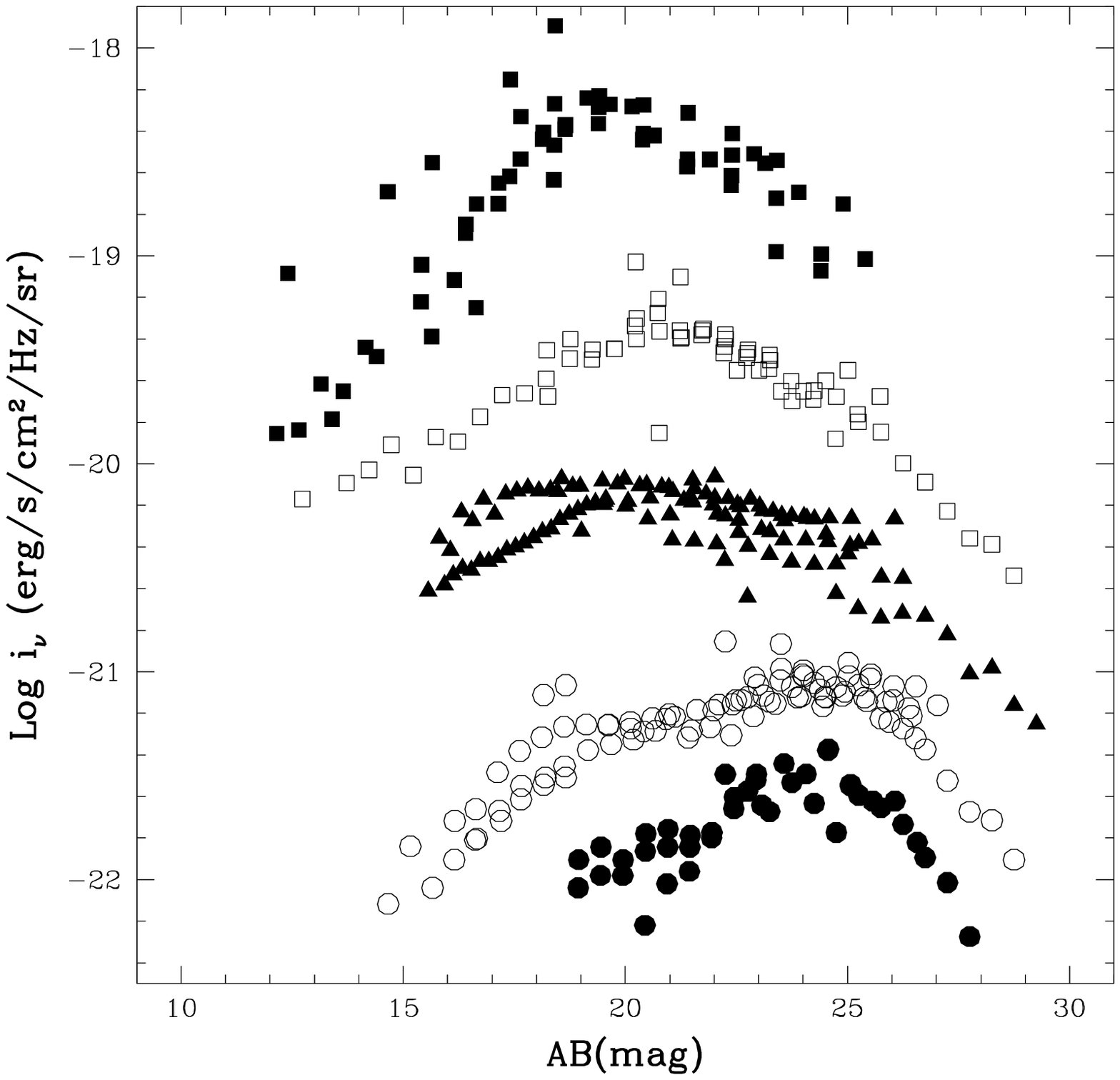}{8}{0}
\caption{{\bf Figure 8.} Extragalactic background light per magnitude bin 
as a function of $U$ ({\it
filled circles}), $B$ ({\it open circles}), $V$ ({\it filled triangles}), $I$
({\it open squares}) and $K$ ({\it filled squares}) magnitudes. For clarity, 
the $B$, $V$, $I$ and $K$ values have been multiplied by a factor of 2, 8,
30, and 100, respectively.
}
\endfigure

We conclude that the results of our analyses, 
i.e. the fluctuation analysis on the $\U3$ data, the colour selection of
Lyman-break galaxies, the derivation of the $\uv$ effective colour as a function
of $\V6$ magnitude (and its application to the $\V6$ counts to predict the
expected ultraviolet counts), are all consistent with each other and 
with the idea that, at faint UV flux limits, there is a significant 
absorption-induced loss of sources in the HDF.
     
\section{Extragalactic Background Light}

The extragalactic background light (EBL) 
is an indicator of the total optical luminosity of the universe. It can
provide unique information on the origin of structures at early epochs, as the
cumulative emission from pregalactic, protogalactic, and evolving galactic
systems is expected to be recorded in this background (Tyson 1995). The
contribution to the EBL from discrete objects can be calculated directly by
integrating the flux times the differential number counts down to the detection
threshold. We have used the compilation of ground-based and HDF data shown in
Figure 2, and the compilation of $K$-band data shown in Figure 3 of 
Pozzetti \etal (1996), to compute the EBL at 
$3500\lta \lambda\lta 22000$\AA. The results are listed in Table 2, along 
with the magnitude range of integration and the estimated 1$\sigma$ error
bars, which arise mostly from field-to-field variations in the numbers of 
relatively bright galaxies.

\fign\beginfigure{\fignumber}
\putfigl{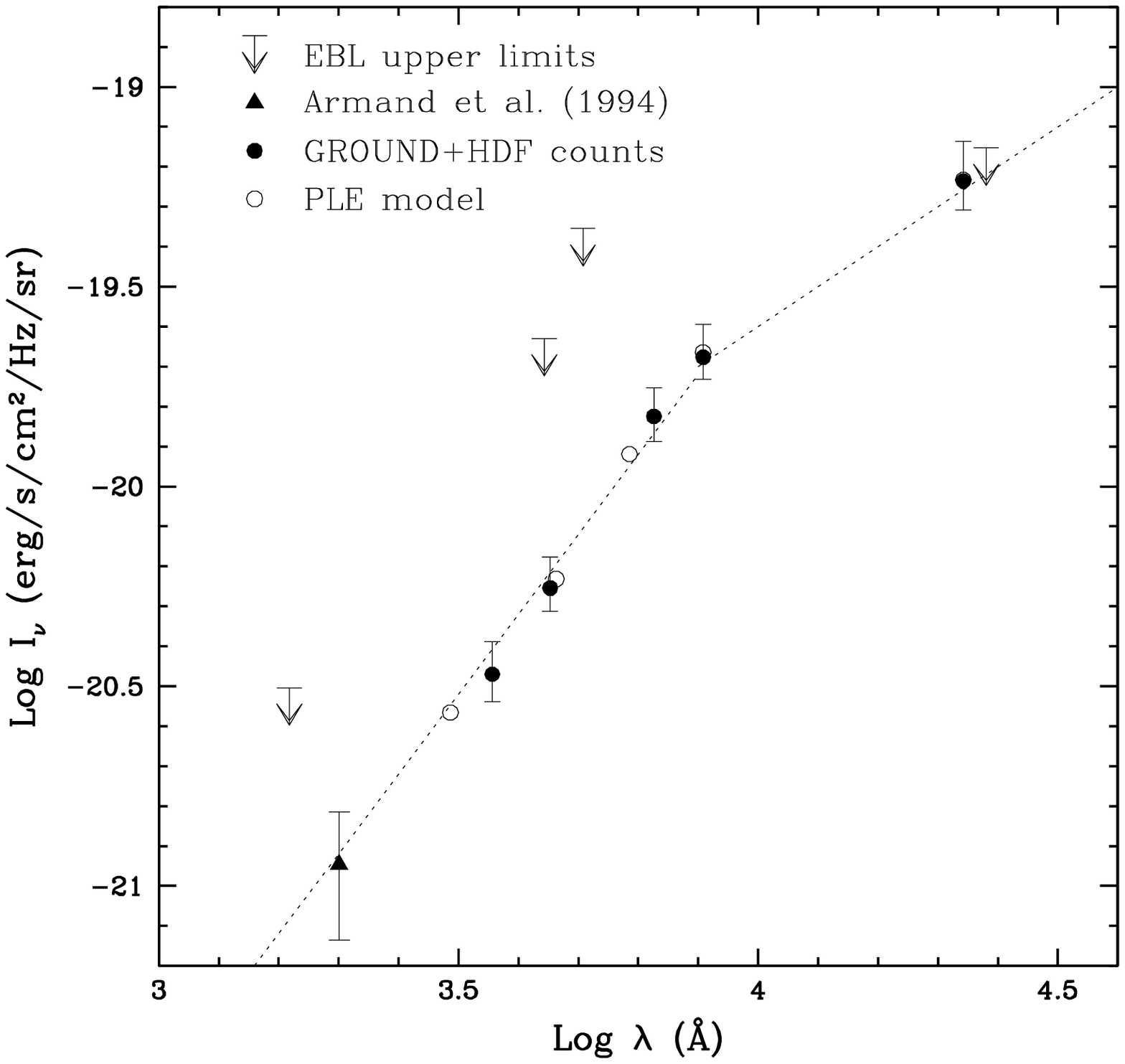}{8}{0}
\caption{{\bf Figure 9.} Spectrum of the extragalactic background light  
as derived from a compilation of ground-based and space-based galaxy counts.
Also plotted are the $90\%$ all-sky-photometry upper limits from Bowyer (1991),
Toller (1983), Dube \etal
(1977, 1979), and Boughn \etal (1986). For clarity, the upper limit at
$\lambda=2.2\mu$m has been plotted at $2.4\mu$m. {\it Dotted line:}
broken power-law fit to the EBL spectrum.
} 
\endfigure

Because of the flattening at faint magnitudes of the $N(m)$ differential 
counts, most of the contribution to the EBL comes from relatively bright 
galaxies.  This is clearly seen in Figure 8, where the function 
$i_\nu=10^{-0.4(m+48.6)}\times N(m)$ is plotted against apparent magnitude in
all bands. While the differential contribution to the EBL increases at bright 
fluxes, where the slope of the ground-based counts is relatively steep, 
it appears that the HDF survey, reaching the limits beyond which the counts 
flatten below $0.40$, has achieved the sensitivity to capture the bulk of 
the extragalactic
light from discrete sources. An extrapolation of the observed $N(m)$
to brighter and/or fainter magnitudes would typically increase the integrated
light by less than 20\%. The spectrum of the EBL (including a UV point at
2000 \AA\ from Armand, Milliard, \& Deharveng 1994) is shown in Figure 9, and
can be  well described by a broken power-law:
$I_\nu=1.9\times 10^{-20} (\lambda/8000\,$\AA$)^{2}
\iunits$ from 2000 to 8000$\,$\AA, and
$I_\nu=1.9\times 10^{-20} (\lambda/8000\,$\AA$) \iunits$ 
from $8000$ to $22000$ \AA.
Several diffuse EBL upper limits -- set by all-sky photometry -- are
also plotted. While in the optical/UV these are from 3 to 5 times higher than
the contribution from known galaxies, the upper limit to the total diffuse
$K$-band EBL lies only slightly above the summed flux from discrete objects. 

\section{Comparison between the Data and a Standard PLE Model}

Qualitatively, the effects of the inclusion of intergalactic attenuation
in any galaxy evolution model are easily described. By severely obscuring 
sources above $z\approx 2$ in the ultraviolet band, and above $z\approx 4$ 
in the blue, the
ubiquitous presence of \HI along the line of sight to distant sources gives
effectively origin to galaxy samples that are volume-limited in these bands. 
This truncates the galaxy redshift distribution and causes a flattening,
more pronounced in the bluest filters, of the slope of the number-magnitude
relation. The differential effect in the various bandpasses could then, at 
least in principle, allow us to discriminate between different models.  
Indeed, a more pronounced flattening of the ultraviolet $N(m)$ relative to  
the counts at longer wavelengths is likely to flag the presence of a 
significant population of faint high-$z$ objects. Moreover, objects close to 
the redshift limit set by the opacity of the intergalactic 
medium should appear quite redder. 

\fign\beginfigure{\fignumber}
\putfigl{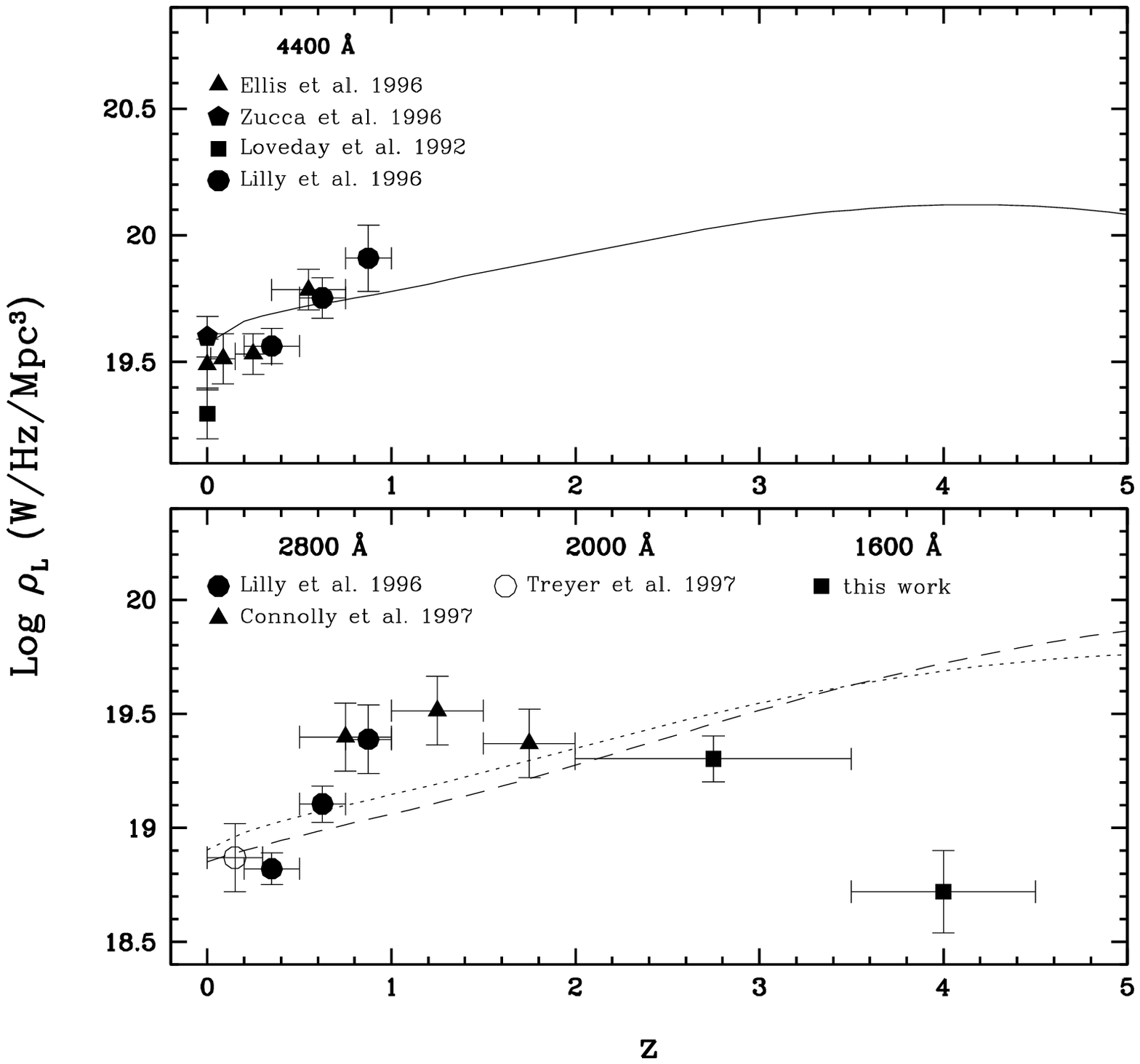}{9}{0}
\caption{{\bf Figure 10.} 
Integrated luminosity per unit comoving volume as a function of redshift
($H_0=50$ km s$^{-1}$ Mpc$^{-1}$, $q_0=0.05$).
The sources of the data points are indicated in each panel. 
The curves show the luminosity density predicted by a PLE model at
4400 \AA\ ({\it solid line}), 2800 \AA\ ({\it dotted line}), and 
1600 \AA\ ({\it dashed line}).
}
\endfigure

In the following, we shall briefly examine 
the predictions of a standard pure-luminosity-evolution
model, modified by the inclusion of the effects of intergalactic attenuation,
and compare them with the results obtained from the HDF data. 
Such a model, which assumes a single redshift of formation ($z_f$)
for all galaxies and a open universe, was shown by Pozzetti \etal (1996) to be 
able to reproduce, with different degrees of accuracy, most of the properties 
of faint field galaxies as derived from ground-based surveys. 
The differential galaxy counts as a functions of AB 
magnitude, as expected within a PLE model with $z_f=6.3$, are shown in 
Figure 2. 
Intervening absorption begins to affect the ultraviolet counts
at $U_{300}\gta 25$ mag, where the predicted slope flattens to  $\gamma\lta 
0.3$. By $U_{300}\sim 29$, the PLE model including intergalactic 
attenuation has 
$\gamma\sim 0$ and predicts about 2 times less galaxies than expected in 
a transparent universe. The median redshift for the global galaxy distribution
is $z_{med}\simeq 2$ at ($U_{300}$, $B_{450}$, $V_{606}$, $I_{814}$) $\simeq$
($29$, $28$, $27$, $26$).  At intermediate magnitudes ($20\lta AB\lta 25$), 
luminosity evolution causes a blueing trend in the mean colours (Pozzetti 
\etal 1996. Only at fainter flux levels ($\gta 25$ mag) the effect of the 
IGM becomes significant, producing a clear trend towards redder 
ultraviolet-optical colours with increasing magnitude (Figure 6).

Overall, the adopted PLE model appears to provide a rather acceptable 
fit to both 
the ground-based and HDF galaxy counts (Figure 2), as well as to the EBL 
spectrum (Figure 9). It produces, in good agreement 
with the observations, a  flattening at faint magnitudes which is more
significant at shorter wavelengths. The predicted turn-over in the
$U_{300}$ bandpass is, however, sharper than allowed by the fluctuation 
analysis, due to the presence of too many high-$z$ galaxies in the model 
relative to data.
While, as shown in Figure 6, the inferred reddening in $\uve$ at magnitudes 
fainter than $V_{606}=26$ is consistent with what expected in a PLE  
model (note how the same model without attenuation predicts bluer colours than 
observed in this magnitude range), at brighter flux levels ($23<V_{606}<26$) 
the synthetic colours appear to be too blue.  The reason for this difference 
between model and data is not clear.

In Figure 4, the predicted fractions of colour-selected $U_{300}$ and $B_{450}$
dropouts versus $V_{606}$ magnitude are compared with the HDF observations.
While the relative numbers of ultraviolet dropouts are 
consistent with the PLE prediction, there is a clear deficit of observed 
$B_{450}$ dropouts relative to the model. As already pointed out by 
Ferguson \& Babul (1997), it appears that a major problem of a PLE 
scenario is the large number of predicted galaxies at high redshifts.  

This is confirmed by Figure 10, which shows a comparison between the observed 
luminosity densities as a function of redshift at different wavelengths and the
predictions of the PLE model. At $4400$ \AA\ the model generates a 
luminosity density which increases only slightly with redshift for $z<1$. The 
local comoving emissivities derived from the data by different authors show 
a large spread, of order a factor two, although the most recent observations 
(Ellis \etal 1996; Zucca \etal 1997) suggest a higher normalization than 
that obtained by Loveday \etal (1992). If this is correct, the data are 
consistent with no or weak evolution up to $z\sim 0.3$. A rapid evolution, 
not reproduced by the model, is instead required by the data in the  
range $0.3 < z < 1.0$, both in the blue and at ultraviolet wavelengths. 
At higher redshifts, the model is in good agreement with the luminosity
density we have derived at $1600$ \AA\ in the range $2.0 < z < 3.5$, while it 
is higher than the observations by a factor $\sim 10$ in the interval 
$3.5<z<4.5$. 

\section {Conclusions}

By applying a simple fluctuation analysis to  the $U_{300}$ data, we have
obtained an estimate of the slope of the ultraviolet counts, down to
more than a  magnitude fainter than the last data point published by 
Williams et al. (1996). The derived UV number-magnitude relation agrees with  
what has been obtained by convolving the observed
$V_{606}$ counts with the derived $(U_{300} - V_{606})_{eff}$ vs. $V_{606}$
relation, and assuming a fraction of high-$z$ galaxies consistent with the 
results of the dropout colour technique. All these results
convincingly show that the effects of an IGM absorption-induced loss of sources
are significant at faint ultraviolet magnitudes ($AB>27$).  
At flux levels of $AB\approx 27$ mag, about 7\% of the sources in $U_{300}$, 
30\% in $B_{450}$, and  35\% in $V_{606}$ are
Lyman-break ``dropouts'', i.e., candidate star-forming galaxies at $z>2$.

The rest-frame UV luminosity densities we derive at high redshifts using
a maximum likelihood reconstruction of the true colour distribution
are in good agreement with previous estimates (Madau et al. 1996; Madau 1997).
We note that the existing data do not necessarily require a sharp peak
of the ultraviolet luminosity density at $z\sim 1.5$, but may also be
consistent with an approximately constant value over a redshift range
as large as $0.8 \lta z \lta 3.0$. 

From our dropout sample we have also derived 
the $M_{1600}$ luminosity functions in the redshift intervals 
$2.0 < z < 3.5$ and $3.5 < z < 4.5$. The LF is reasonably well defined 
in the lower redshift interval, and can 
be described by a Schechter function with parameters $\alpha=-1.3$,
$M_{1600}^*=-21.7$ and $\Phi^*=0.8\times 10^{-3}$ Mpc$^{-3}$.
The same $\alpha$ and $M^*_{1600}$ values, but with a normalization four times 
lower, provide a good fit to the blue dropouts at higher redshifts. Both 
functions extend to brighter magnitudes than the local 2000 \AA\ LF  
recently derived by Treyer et al. (1997).

A detailed comparison between a simple PLE galaxy evolution model 
and the available ground-based and HDF data shows that the model 
fails to reproduce all the observed properties of faint field galaxies.
While able to fit reasonably well the observed number counts, 
colours, and the fraction of ultraviolet dropouts at
$\langle z \rangle \sim 2.75$, this PLE model overpredicts by about one order
of magnitude the number of blue dropouts at $z>3.5$ and the corresponding 
rest-frame UV luminosity density. Moreover, it 
fails to account for the steep observed trend of the galaxy emissivity 
observed in the redshift range $0.3<z<1$ at all wavelengths. 
The modeling of the emission history of the galaxy population at late epochs
is outside the scope of this paper; we note
here that Madau, Pozzetti \& Dickinson (1997) have recently shown that the 
rapid variation of the luminosity density at small redshifts can be 
well fit by a simple stellar evolution model, defined by  time-dependent 
star formation rate per unit comoving volume and a universal initial mass
function.  At high redshifts, since in our PLE model a significant fraction 
of the emitted light is due to elliptical galaxies just after their first 
massive starburst phase, a possible solution to the observed discrepancy 
could be obtained by assuming that a significant amount of dust obscures
the early evolution of these massive systems. A similar conclusion has been
reached by Franceschini et al (1997) in their analysis of a K-band selected
sample of early-type galaxies in the HDF. Approximate consistency with the 
$\langle z \rangle \sim 4 $ observations would require a dust extinction 
of $E(B-V)\sim 0.2\div0.23$ for SMC--type dust and the empirical 
extinction--law of Calzetti, Kinney \& Storchi-Bergman (1994), respectively,
corresponding to an ultraviolet extinction of about a factor of $10$. 
The current estimates of the ultraviolet extinction for the star-forming
galaxies at $z \sim 3$ are still very uncertain and range from about 2
(Pettini et al. 1997) to more than 10 (Meurer \etal 1997).
If such a large amount of dust, consistent with the observations of local
starburst galaxies (Calzetti 1997), was present in star-forming objects 
at high-$z$, it would absorb a large fraction of the optical--UV light 
and re-emit it in the far infrared band (see, for example, the models 
described in detail in Mazzei, De Zotti \& Xu 1994). In this case, the
amount of star formation derived from our and similar dropout analyses
should only be considered as a lower limit. Preliminary observations of the 
HDF with ISO (Rowan-Robinson et al. 1997) have revealed about a dozen 
far-IR ($6.7$ and $15\mu m$) sources, identified with optical galaxies. 
By fitting the spectra of these galaxies from the ultraviolet to the far-IR,
Rowan-Robinson et al. (1997) have derived star formation rates significantly 
higher than those derived from the UV fluxes detected with HST. On the other 
hand, Madau \etal (1997) show that a model with a large amount of hidden star 
formation at early epochs will overproduce the cosmic metallicity at 
high-$z$ as sampled by the damped Lyman-$\alpha$  systems (Pettini et al. 1997).
The results discussed above demonstrate the great 
potential of IR observations for complementing the information provided by 
the rest-frame ultraviolet  and for helping our understanding
of the nature of high-$z$ galaxies.
 
\section*{Acknowledgments}

We have benefited from discussions with M. Dickinson, A. Fruchter, and 
M. Romaniello. Support for this work was provided by NASA through grant
AR-06337.10-94A from the Space Telescope Science Institute, which is operated
by the Association of Universities for Research in Astronomy, Inc., under NASA
contract NAS5-26555, and by ASI through contracts 95-RS-152 and ARS-96-70.

\bye
\end

%% file: mn.tex
%
%
%
%

\catcode `\@=11 

\def\@version{1.4}
\def\@verdate{22nd Feb 1994}

%
%
%
%


\newif\ifprod@font

\ifx\@typeface\undefined
  \def\@typeface{Comp. Modern}\prod@fontfalse
\else
  \prod@fonttrue 
\fi

\def\newfam{\alloc@8\fam\chardef\sixt@@n} 

\ifprod@font
\font\fiverm=mtr10 at 5pt
\font\fivebf=mtbx10 at 5pt
\font\fiveit=mtti10 at 5pt
\font\fivesl=mtsl10 at 5pt
\font\fivett=mttt10 at 5pt     \hyphenchar\fivett=-1
\font\fivecsc=mtcsc10 at 5pt
\font\fivesf=mtss10 at 5pt
\font\fivei=mtmi10 at 5pt      \skewchar\fivei='177
\font\fivemib=mtmib10 at 5pt   \skewchar\fivemib='177
\font\fivesy=mtsy10 at 5pt     \skewchar\fivesy='60
\font\fivesyb=mtbsy10 at 5pt   \skewchar\fivesyb='60

\font\sixrm=mtr10 at 6pt
\font\sixbf=mtbx10 at 6pt
\font\sixit=mtti10 at 6pt
\font\sixsl=mtsl10 at 6pt
\font\sixtt=mttt10 at 6pt      \hyphenchar\sixtt=-1
\font\sixcsc=mtcsc10 at 6pt
\font\sixsf=mtss10 at 6pt
\font\sixi=mtmi10 at 6pt       \skewchar\sixi='177
\font\sixmib=mtmib10 at 6pt    \skewchar\sixmib='177
\font\sixsy=mtsy10 at 6pt      \skewchar\sixsy='60
\font\sixsyb=mtbsy10 at 6pt    \skewchar\sixsyb='60

\font\sevenrm=mtr10 at 7pt
\font\sevenbf=mtbx10 at 7pt
\font\sevenit=mtti10 at 7pt
\font\sevensl=mtsl10 at 7pt
\font\seventt=mttt10 at 7pt     \hyphenchar\seventt=-1
\font\sevencsc=mtcsc10 at 7pt
\font\sevensf=mtss10 at 7pt
\font\seveni=mtmi10 at 7pt      \skewchar\seveni='177
\font\sevenmib=mtmib10 at 7pt   \skewchar\sevenmib='177
\font\sevensy=mtsy10 at 7pt     \skewchar\sevensy='60
\font\sevensyb=mtbsy10 at 7pt   \skewchar\sevensyb='60

\font\eightrm=mtr10 at 8pt
\font\eightbf=mtbx10 at 8pt
\font\eightit=mtti10 at 8pt
\font\eighti=mtmi10 at 8pt      \skewchar\eighti='177
\font\eightmib=mtmib10 at 8pt   \skewchar\eightmib='177
\font\eightsy=mtsy10 at 8pt     \skewchar\eightsy='60
\font\eightsyb=mtbsy10 at 8pt   \skewchar\eightsyb='60
\font\eightsl=mtsl10 at 8pt
\font\eighttt=mttt10 at 8pt     \hyphenchar\eighttt=-1
\font\eightcsc=mtcsc10 at 8pt
\font\eightsf=mtss10 at 8pt

\font\ninerm=mtr10 at 9pt
\font\ninebf=mtbx10 at 9pt
\font\nineit=mtti10 at 9pt
\font\ninei=mtmi10 at 9pt      \skewchar\ninei='177
\font\ninemib=mtmib10 at 9pt   \skewchar\ninemib='177
\font\ninesy=mtsy10 at 9pt     \skewchar\ninesy='60
\font\ninesyb=mtbsy10 at 9pt   \skewchar\ninesyb='60
\font\ninesl=mtsl10 at 9pt
\font\ninett=mttt10 at 9pt     \hyphenchar\ninett=-1
\font\ninecsc=mtcsc10 at 9pt
\font\ninesf=mtss10 at 9pt

\font\tenrm=mtr10
\font\tenbf=mtbx10
\font\tenit=mtti10
\font\teni=mtmi10		\skewchar\teni='177
\font\tenmib=mtmib10	\skewchar\tenmib='177
\font\tensy=mtsy10		\skewchar\tensy='60
\font\tensyb=mtbsy10	\skewchar\tensyb='60
\font\tenex=cmex10
\font\tensl=mtsl10
\font\tentt=mttt10		\hyphenchar\tentt=-1
\font\tencsc=mtcsc10
\font\tensf=mtss10

\font\elevenrm=mtr10 at 11pt
\font\elevenbf=mtbx10 at 11pt
\font\elevenit=mtti10 at 11pt
\font\eleveni=mtmi10 at 11pt      \skewchar\eleveni='177
\font\elevenmib=mtmib10 at 11pt   \skewchar\elevenmib='177
\font\elevensy=mtsy10 at 11pt     \skewchar\elevensy='60
\font\elevensyb=mtbsy10 at 11pt   \skewchar\elevensyb='60
\font\elevensl=mtsl10 at 11pt
\font\eleventt=mttt10 at 11pt     \hyphenchar\eleventt=-1
\font\elevencsc=mtcsc10 at 11pt
\font\elevensf=mtss10 at 11pt

\font\twelverm=mtr10 at 12pt
\font\twelvebf=mtbx10 at 12pt
\font\twelveit=mtti10 at 12pt
\font\twelvesl=mtsl10 at 12pt
\font\twelvett=mttt10 at 12pt     \hyphenchar\twelvett=-1
\font\twelvecsc=mtcsc10 at 12pt
\font\twelvesf=mtss10 at 12pt
\font\twelvei=mtmi10 at 12pt      \skewchar\twelvei='177
\font\twelvemib=mtmib10 at 12pt   \skewchar\twelvemib='177
\font\twelvesy=mtsy10 at 12pt     \skewchar\twelvesy='60
\font\twelvesyb=mtbsy10 at 12pt   \skewchar\twelvesyb='60

\font\fourteenrm=mtr10 at 14pt
\font\fourteenbf=mtbx10 at 14pt
\font\fourteenit=mtti10 at 14pt
\font\fourteeni=mtmi10 at 14pt      \skewchar\fourteeni='177
\font\fourteenmib=mtmib10 at 14pt   \skewchar\fourteenmib='177
\font\fourteensy=mtsy10 at 14pt     \skewchar\fourteensy='60
\font\fourteensyb=mtbsy10 at 14pt   \skewchar\fourteensyb='60
\font\fourteensl=mtsl10 at 14pt
\font\fourteentt=mttt10 at 14pt     \hyphenchar\fourteentt=-1
\font\fourteencsc=mtcsc10 at 14pt
\font\fourteensf=mtss10 at 14pt

\font\seventeenrm=mtr10 at 17pt
\font\seventeenbf=mtbx10 at 17pt
\font\seventeenit=mtti10 at 17pt
\font\seventeeni=mtmi10 at 17pt      \skewchar\seventeeni='177
\font\seventeenmib=mtmib10 at 17pt   \skewchar\seventeenmib='177
\font\seventeensy=mtsy10 at 17pt     \skewchar\seventeensy='60
\font\seventeensyb=mtbsy10 at 17pt   \skewchar\seventeensyb='60
\font\seventeensl=mtsl10 at 17pt
\font\seventeentt=mttt10 at 17pt     \hyphenchar\seventeentt=-1
\font\seventeencsc=mtcsc10 at 17pt
\font\seventeensf=mtss10 at 17pt


\newfam\xmfam
\newfam\ymfam

\font\fivexm=mtxm10 at 5pt
\font\sixxm=mtxm10 at 6pt
\font\sevenxm=mtxm10 at 7pt
\font\eightxm=mtxm10 at 8pt
\font\ninexm=mtxm10 at 9pt
\font\tenxm=mtxm10
\font\elevenxm=mtxm10 at 11pt
\font\twelvexm=mtxm10 at 12pt
\font\fourteenxm=mtxm10 at 14pt
\font\seventeenxm=mtxm10 at 17pt

\font\fiveym=mtym10 at 5pt
\font\sixym=mtym10 at 6pt
\font\sevenym=mtym10 at 7pt
\font\eightym=mtym10 at 8pt
\font\nineym=mtym10 at 9pt
\font\tenym=mtym10
\font\elevenym=mtym10 at 11pt
\font\twelveym=mtym10 at 12pt
\font\fourteenym=mtym10 at 14pt
\font\seventeenym=mtym10 at 17pt
\else
\font\fiverm=cmr5
\font\fivei=cmmi5             \skewchar\fivei='177
\font\fivemib=cmmib10 at 5pt  \skewchar\fivemib='177
\font\fivesy=cmsy5            \skewchar\fivesy='60
\font\fivesyb=cmbsy10 at 5pt  \skewchar\fivesyb='60
\font\fivebf=cmbx5

\font\sixrm=cmr6
\font\sixi=cmmi6             \skewchar\sixi='177
\font\sixmib=cmmib10 at 6pt  \skewchar\sixmib='177
\font\sixsy=cmsy6            \skewchar\sixsy='60
\font\sixsyb=cmbsy10 at 6pt  \skewchar\sixsyb='60
\font\sixbf=cmbx6

\font\sevenrm=cmr7
\font\seveni=cmmi7             \skewchar\seveni='177
\font\sevenmib=cmmib10 at 7pt  \skewchar\sevenmib='177
\font\sevensy=cmsy7            \skewchar\sevensy='60
\font\sevensyb=cmbsy10 at 7pt  \skewchar\sevensyb='60
\font\sevenbf=cmbx7

\font\eightrm=cmr8
\font\eightbf=cmbx8
\font\eightit=cmti8
\font\eighti=cmmi8			\skewchar\eighti='177
\font\eightmib=cmmib10 at 8pt	\skewchar\eightmib='177
\font\eightsy=cmsy8			\skewchar\eightsy='60
\font\eightsyb=cmbsy10 at 8pt	\skewchar\eightsyb='60
\font\eightsl=cmsl8
\font\eighttt=cmtt8			\hyphenchar\eighttt=-1
\font\eightcsc=cmcsc10 at 8pt
\font\eightsf=cmss8

\font\ninerm=cmr9
\font\ninebf=cmbx9
\font\nineit=cmti9
\font\ninei=cmmi9			\skewchar\ninei='177
\font\ninemib=cmmib10 at 9pt	\skewchar\ninemib='177
\font\ninesy=cmsy9			\skewchar\ninesy='60
\font\ninesyb=cmbsy10 at 9pt	\skewchar\ninesyb='60
\font\ninesl=cmsl9
\font\ninett=cmtt9			\hyphenchar\ninett=-1
\font\ninecsc=cmcsc10 at 9pt
\font\ninesf=cmss9

\font\tenrm=cmr10
\font\tenbf=cmbx10
\font\tenit=cmti10
\font\teni=cmmi10		\skewchar\teni='177
\font\tenmib=cmmib10	\skewchar\tenmib='177
\font\tensy=cmsy10		\skewchar\tensy='60
\font\tensyb=cmbsy10	\skewchar\tensyb='60
\font\tenex=cmex10
\font\tensl=cmsl10
\font\tentt=cmtt10		\hyphenchar\tentt=-1
\font\tencsc=cmcsc10
\font\tensf=cmss10

\font\elevenrm=cmr10 scaled \magstephalf
\font\elevenbf=cmbx10 scaled \magstephalf
\font\elevenit=cmti10 scaled \magstephalf
\font\eleveni=cmmi10 scaled \magstephalf	\skewchar\eleveni='177
\font\elevenmib=cmmib10 scaled \magstephalf	\skewchar\elevenmib='177
\font\elevensy=cmsy10 scaled \magstephalf	\skewchar\elevensy='60
\font\elevensyb=cmbsy10 scaled \magstephalf	\skewchar\elevensyb='60
\font\elevensl=cmsl10 scaled \magstephalf
\font\eleventt=cmtt10 scaled \magstephalf	\hyphenchar\eleventt=-1
\font\elevencsc=cmcsc10 scaled \magstephalf
\font\elevensf=cmss10 scaled \magstephalf

\font\twelverm=cmr10 scaled \magstep1
\font\twelvebf=cmbx10 scaled \magstep1
\font\twelvei=cmmi10 scaled \magstep1      \skewchar\twelvei='177
\font\twelvemib=cmmib10 scaled \magstep1   \skewchar\twelvemib='177
\font\twelvesy=cmsy10 scaled \magstep1     \skewchar\twelvesy='60
\font\twelvesyb=cmbsy10 scaled \magstep1   \skewchar\twelvesyb='60

\font\fourteenrm=cmr10 scaled \magstep2
\font\fourteenbf=cmbx10 scaled \magstep2
\font\fourteenit=cmti10 scaled \magstep2
\font\fourteeni=cmmi10 scaled \magstep2		\skewchar\fourteeni='177
\font\fourteenmib=cmmib10 scaled \magstep2	\skewchar\fourteenmib='177
\font\fourteensy=cmsy10 scaled \magstep2	\skewchar\fourteensy='60
\font\fourteensyb=cmbsy10 scaled \magstep2	\skewchar\fourteensyb='60
\font\fourteensl=cmsl10 scaled \magstep2
\font\fourteentt=cmtt10 scaled \magstep2	\hyphenchar\fourteentt=-1
\font\fourteencsc=cmcsc10 scaled \magstep2
\font\fourteensf=cmss10 scaled \magstep2

\font\seventeenrm=cmr10 scaled \magstep3
\font\seventeenbf=cmbx10 scaled \magstep3
\font\seventeenit=cmti10 scaled \magstep3
\font\seventeeni=cmmi10 scaled \magstep3	\skewchar\seventeeni='177
\font\seventeenmib=cmmib10 scaled \magstep3	\skewchar\seventeenmib='177
\font\seventeensy=cmsy10 scaled \magstep3	\skewchar\seventeensy='60
\font\seventeensyb=cmbsy10 scaled \magstep3	\skewchar\seventeensyb='60
\font\seventeensl=cmsl10 scaled \magstep3
\font\seventeentt=cmtt10 scaled \magstep3	\hyphenchar\seventeentt=-1
\font\seventeencsc=cmcsc10 scaled \magstep3
\font\seventeensf=cmss10 scaled \magstep3
\fi

\def\hexnumber#1{\ifcase#1 0\or1\or2\or3\or4\or5\or6\or7\or8\or9\or
  A\or B\or C\or D\or E\or F\fi}

\ifprod@font
  \edef\@xm{\hexnumber\xmfam}
  \edef\@ym{\hexnumber\ymfam}
\fi

\def\makestrut{%
  \setbox\strutbox=\hbox{%
    \vrule height.7\baselineskip depth.3\baselineskip width \z@}%
}

\def\baselinestretch{1}
\newskip\tmp@bls

\def\b@ls#1{
  \tmp@bls=#1\relax
  \baselineskip=#1\relax\makestrut
  \normalbaselineskip=\baselinestretch\tmp@bls
  \normalbaselines
}

\def\nostb@ls#1{
  \normalbaselineskip=#1\relax
  \normalbaselines
  \makestrut
}

%

\newfam\mibfam 
\newfam\sybfam 
\newfam\scfam  
\newfam\sffam  

\def\mit{\fam\@ne}

\def\cal{\fam\tw@}

\def\em{\ifdim\fontdimen1\font>\z@ \rm\else\it\fi}

\textfont3=\tenex
\scriptfont3=\tenex
\scriptscriptfont3=\tenex

\setbox0=\hbox{\tenex B} \p@renwd=\wd0 

\def\eightpoint{
  \def\rm{\fam0\eightrm}%
  \textfont0=\eightrm \scriptfont0=\sixrm \scriptscriptfont0=\fiverm%
  \textfont1=\eighti  \scriptfont1=\sixi  \scriptscriptfont1=\fivei%
  \textfont2=\eightsy \scriptfont2=\sixsy \scriptscriptfont2=\fivesy%
  \textfont\itfam=\eightit\def\it{\fam\itfam\eightit}%
  \ifprod@font
    \scriptfont\itfam=\sixit
      \scriptscriptfont\itfam=\fiveit
  \else
    \scriptfont\itfam=\eightit
      \scriptscriptfont\itfam=\eightit
  \fi
  \textfont\bffam=\eightbf%
    \scriptfont\bffam=\sixbf%
      \scriptscriptfont\bffam=\fivebf%
  \def\bf{\fam\bffam\eightbf}%
  \textfont\slfam=\eightsl\def\sl{\fam\slfam\eightsl}%
  \ifprod@font
    \scriptfont\slfam=\sixsl
      \scriptscriptfont\slfam=\fivesl
  \else
    \scriptfont\slfam=\eightsl
      \scriptscriptfont\slfam=\eightsl
  \fi
  \textfont\ttfam=\eighttt\def\tt{\fam\ttfam\eighttt}%
  \ifprod@font
    \scriptfont\ttfam=\sixtt
      \scriptscriptfont\ttfam=\fivett
  \else
    \scriptfont\ttfam=\eighttt
      \scriptscriptfont\ttfam=\eighttt
  \fi
  \textfont\scfam=\eightcsc\def\sc{\fam\scfam\eightcsc}%
  \ifprod@font
    \scriptfont\scfam=\sixcsc
      \scriptscriptfont\scfam=\fivecsc
  \else
    \scriptfont\scfam=\eightcsc
      \scriptscriptfont\scfam=\eightcsc
  \fi
  \textfont\sffam=\eightsf\def\sf{\fam\sffam\eightsf}%
  \ifprod@font
    \scriptfont\sffam=\sixsf
      \scriptscriptfont\sffam=\fivesf
  \else
    \scriptfont\sffam=\eightsf
      \scriptscriptfont\sffam=\eightsf
  \fi
  \textfont\mibfam=\eightmib
    \scriptfont\mibfam=\sixmib
      \scriptscriptfont\mibfam=\fivemib
  \textfont\sybfam=\eightsyb
    \scriptfont\sybfam=\sixsyb
      \scriptscriptfont\sybfam=\fivesyb
  \ifprod@font
    \textfont\xmfam=\eightxm
      \scriptfont\xmfam=\sixxm
        \scriptscriptfont\xmfam=\fivexm
    \textfont\ymfam=\eightym
      \scriptfont\ymfam=\sixym
        \scriptscriptfont\ymfam=\fiveym
  \fi
  \def\oldstyle{\fam\@ne\eighti}%
  \def\boldstyle{\fam\mibfam\eightmib}%
  \b@ls{10pt}\rm%
}

\def\ninepoint{
  \def\rm{\fam0\ninerm}%
  \textfont0=\ninerm \scriptfont0=\sixrm \scriptscriptfont0=\fiverm%
  \textfont1=\ninei  \scriptfont1=\sixi  \scriptscriptfont1=\fivei%
  \textfont2=\ninesy \scriptfont2=\sixsy \scriptscriptfont2=\fivesy%
  \textfont\itfam=\nineit\def\it{\fam\itfam\nineit}%
  \ifprod@font
    \scriptfont\itfam=\sixit
      \scriptscriptfont\itfam=\fiveit
  \else
    \scriptfont\itfam=\nineit
      \scriptscriptfont\itfam=\nineit
  \fi
  \textfont\bffam=\ninebf%
    \scriptfont\bffam=\sixbf%
      \scriptscriptfont\bffam=\fivebf%
  \def\bf{\fam\bffam\ninebf}%
  \textfont\slfam=\ninesl\def\sl{\fam\slfam\ninesl}%
  \ifprod@font
    \scriptfont\slfam=\sixsl
      \scriptscriptfont\slfam=\fivesl
  \else
    \scriptfont\slfam=\ninesl
      \scriptscriptfont\slfam=\ninesl
  \fi
  \textfont\ttfam=\ninett\def\tt{\fam\ttfam\ninett}%
  \ifprod@font
    \scriptfont\ttfam=\sixtt
      \scriptscriptfont\ttfam=\fivett
  \else
    \scriptfont\ttfam=\ninett
      \scriptscriptfont\ttfam=\ninett
  \fi
  \textfont\scfam=\ninecsc\def\sc{\fam\scfam\ninecsc}%
  \ifprod@font
    \scriptfont\scfam=\sixcsc
      \scriptscriptfont\scfam=\fivecsc
  \else
    \scriptfont\scfam=\ninecsc
      \scriptscriptfont\scfam=\ninecsc
  \fi
  \textfont\sffam=\ninesf\def\sf{\fam\sffam\ninesf}%
  \ifprod@font
    \scriptfont\sffam=\sixsf
      \scriptscriptfont\sffam=\fivesf
  \else
    \scriptfont\sffam=\ninesf
      \scriptscriptfont\sffam=\ninesf
  \fi
  \textfont\mibfam=\ninemib
    \scriptfont\mibfam=\sixmib
      \scriptscriptfont\mibfam=\fivemib
  \textfont\sybfam=\ninesyb
    \scriptfont\sybfam=\sixsyb
      \scriptscriptfont\sybfam=\fivesyb
  \ifprod@font
    \textfont\xmfam=\ninexm
      \scriptfont\xmfam=\sixxm
        \scriptscriptfont\xmfam=\fivexm
    \textfont\ymfam=\nineym
      \scriptfont\ymfam=\sixym
        \scriptscriptfont\ymfam=\fiveym
  \fi
  \def\oldstyle{\fam\@ne\ninei}%
  \def\boldstyle{\fam\mibfam\ninemib}%
  \b@ls{\TextLeading plus \Feathering}\rm%
}

\def\tenpoint{
  \def\rm{\fam0\tenrm}%
  \textfont0=\tenrm \scriptfont0=\sevenrm \scriptscriptfont0=\fiverm%
  \textfont1=\teni  \scriptfont1=\seveni  \scriptscriptfont1=\fivei%
  \textfont2=\tensy \scriptfont2=\sevensy \scriptscriptfont2=\fivesy%
  \textfont\itfam=\tenit\def\it{\fam\itfam\tenit}%
  \ifprod@font
    \scriptfont\itfam=\sevenit
      \scriptscriptfont\itfam=\fiveit
  \else
    \scriptfont\itfam=\tenit
      \scriptscriptfont\itfam=\tenit
  \fi
  \textfont\bffam=\tenbf%
    \scriptfont\bffam=\sevenbf%
      \scriptscriptfont\bffam=\fivebf%
  \def\bf{\fam\bffam\tenbf}%
  \textfont\slfam=\tensl\def\sl{\fam\slfam\tensl}%
  \ifprod@font
    \scriptfont\slfam=\sevensl
      \scriptscriptfont\slfam=\fivesl
  \else
    \scriptfont\slfam=\tensl
      \scriptscriptfont\slfam=\tensl
  \fi
  \textfont\ttfam=\tentt\def\tt{\fam\ttfam\tentt}%
  \ifprod@font
    \scriptfont\ttfam=\seventt
      \scriptscriptfont\ttfam=\fivett
  \else
    \scriptfont\ttfam=\tentt
      \scriptscriptfont\ttfam=\tentt
  \fi
  \textfont\scfam=\tencsc\def\sc{\fam\scfam\tencsc}%
  \ifprod@font
    \scriptfont\scfam=\sevencsc
      \scriptscriptfont\scfam=\fivecsc
  \else
    \scriptfont\scfam=\tencsc
      \scriptscriptfont\scfam=\tencsc
  \fi
  \textfont\sffam=\tensf\def\sf{\fam\sffam\tensf}%
  \ifprod@font
    \scriptfont\sffam=\sevensf
      \scriptscriptfont\sffam=\fivesf
  \else
    \scriptfont\sffam=\tensf
      \scriptscriptfont\sffam=\tensf
  \fi
  \textfont\mibfam=\tenmib
    \scriptfont\mibfam=\sevenmib
      \scriptscriptfont\mibfam=\fivemib
  \textfont\sybfam=\tensyb
    \scriptfont\sybfam=\sevensyb
      \scriptscriptfont\sybfam=\fivesyb
  \ifprod@font
    \textfont\xmfam=\tenxm
      \scriptfont\xmfam=\sevenxm
        \scriptscriptfont\xmfam=\fivexm
    \textfont\ymfam=\tenym
      \scriptfont\ymfam=\sevenym
        \scriptscriptfont\ymfam=\fiveym
  \fi
  \def\oldstyle{\fam\@ne\teni}%
  \def\boldstyle{\fam\mibfam\tenmib}%
  \b@ls{11pt}\rm%
}

\def\elevenpoint{
  \def\rm{\fam0\elevenrm}%
  \textfont0=\elevenrm \scriptfont0=\eightrm \scriptscriptfont0=\sixrm%
  \textfont1=\eleveni  \scriptfont1=\eighti  \scriptscriptfont1=\sixi%
  \textfont2=\elevensy \scriptfont2=\eightsy \scriptscriptfont2=\sixsy%
  \textfont\itfam=\elevenit\def\it{\fam\itfam\elevenit}%
  \ifprod@font
    \scriptfont\itfam=\eightit
      \scriptscriptfont\itfam=\sixit
  \else
    \scriptfont\itfam=\elevenit
      \scriptscriptfont\itfam=\elevenit
  \fi
  \textfont\bffam=\elevenbf%
    \scriptfont\bffam=\eightbf%
      \scriptscriptfont\bffam=\sixbf%
  \def\bf{\fam\bffam\elevenbf}%
  \textfont\slfam=\elevensl\def\sl{\fam\slfam\elevensl}%
  \ifprod@font
    \scriptfont\slfam=\eightsl
      \scriptscriptfont\slfam=\sixsl
  \else
    \scriptfont\slfam=\elevensl
      \scriptscriptfont\slfam=\elevensl
  \fi
  \textfont\ttfam=\eleventt\def\tt{\fam\ttfam\eleventt}%
  \ifprod@font
    \scriptfont\ttfam=\eighttt
      \scriptscriptfont\ttfam=\sixtt
  \else
    \scriptfont\ttfam=\eleventt
      \scriptscriptfont\ttfam=\eleventt
  \fi
  \textfont\scfam=\elevencsc\def\sc{\fam\scfam\elevencsc}%
  \ifprod@font
    \scriptfont\scfam=\eightcsc
      \scriptscriptfont\scfam=\sixcsc
  \else
    \scriptfont\scfam=\elevencsc
      \scriptscriptfont\scfam=\elevencsc
  \fi
  \textfont\sffam=\elevensf\def\sf{\fam\sffam\elevensf}%
  \ifprod@font
    \scriptfont\sffam=\eightsf
      \scriptscriptfont\sffam=\sixsf
  \else
    \scriptfont\sffam=\elevensf
      \scriptscriptfont\sffam=\elevensf
  \fi
  \textfont\mibfam=\elevenmib
    \scriptfont\mibfam=\eightmib
      \scriptscriptfont\mibfam=\sixmib
  \textfont\sybfam=\elevensyb
    \scriptfont\sybfam=\eightsyb
      \scriptscriptfont\sybfam=\sixsyb
  \ifprod@font
    \textfont\xmfam=\elevenxm
      \scriptfont\xmfam=\eightxm
       \scriptscriptfont\xmfam=\sixxm
    \textfont\ymfam=\elevenym
      \scriptfont\ymfam=\eightym
        \scriptscriptfont\ymfam=\sixym
   \fi
  \def\oldstyle{\fam\@ne\eleveni}%
  \def\boldstyle{\fam\mibfam\elevenmib}%
  \b@ls{13pt}\rm%
}

\def\fourteenpoint{
  \def\rm{\fam0\fourteenrm}%
  \textfont0\fourteenrm  \scriptfont0\tenrm  \scriptscriptfont0\sevenrm%
  \textfont1\fourteeni   \scriptfont1\teni   \scriptscriptfont1\seveni%
  \textfont2\fourteensy  \scriptfont2\tensy  \scriptscriptfont2\sevensy%
  \textfont\itfam=\fourteenit\def\it{\fam\itfam\fourteenit}%
  \ifprod@font
    \scriptfont\itfam=\tenit
      \scriptscriptfont\itfam=\sevenit
  \else
    \scriptfont\itfam=\fourteenit
      \scriptscriptfont\itfam=\fourteenit
  \fi
  \textfont\bffam=\fourteenbf%
    \scriptfont\bffam=\tenbf%
      \scriptscriptfont\bffam=\sevenbf%
  \def\bf{\fam\bffam\fourteenbf}%
  \textfont\slfam=\fourteensl\def\sl{\fam\slfam\fourteensl}%
  \ifprod@font
    \scriptfont\slfam=\tensl
      \scriptscriptfont\slfam=\sevensl
  \else
    \scriptfont\slfam=\fourteensl
      \scriptscriptfont\slfam=\fourteensl
  \fi
  \textfont\ttfam=\fourteentt\def\tt{\fam\ttfam\fourteentt}%
  \ifprod@font
    \scriptfont\ttfam=\tentt
      \scriptscriptfont\ttfam=\seventt
  \else
    \scriptfont\ttfam=\fourteentt
      \scriptscriptfont\ttfam=\fourteentt
  \fi
  \textfont\scfam=\fourteencsc\def\sc{\fam\scfam\fourteencsc}%
  \ifprod@font
    \scriptfont\scfam=\tencsc
      \scriptscriptfont\scfam=\sevencsc
  \else
    \scriptfont\scfam=\fourteencsc
      \scriptscriptfont\scfam=\fourteencsc
  \fi
  \textfont\sffam=\fourteensf\def\sf{\fam\sffam\fourteensf}%
  \ifprod@font
    \scriptfont\sffam=\tensf
      \scriptscriptfont\sffam=\sevensf
  \else
    \scriptfont\sffam=\fourteensf
      \scriptscriptfont\sffam=\fourteensf
  \fi
  \textfont\mibfam=\fourteenmib
    \scriptfont\mibfam=\tenmib
      \scriptscriptfont\mibfam=\sevenmib
  \textfont\sybfam=\fourteensyb
    \scriptfont\sybfam=\tensyb
      \scriptscriptfont\sybfam=\sevensyb
  \ifprod@font
    \textfont\xmfam=\fourteenxm
      \scriptfont\xmfam=\tenxm
        \scriptscriptfont\xmfam=\sevenxm
   \textfont\ymfam=\fourteenym
      \scriptfont\ymfam=\tenym
        \scriptscriptfont\ymfam=\sevenym
  \fi
  \def\oldstyle{\fam\@ne\fourteeni}%
  \def\boldstyle{\fam\mibfam\fourteenmib}%
  \b@ls{17pt}\rm%
}

\def\seventeenpoint{
  \def\rm{\fam0\seventeenrm}%
  \textfont0\seventeenrm  \scriptfont0\twelverm  \scriptscriptfont0\tenrm%
  \textfont1\seventeeni   \scriptfont1\twelvei   \scriptscriptfont1\teni%
  \textfont2\seventeensy  \scriptfont2\twelvesy  \scriptscriptfont2\tensy%
  \textfont\itfam=\seventeenit\def\it{\fam\itfam\seventeenit}%
  \ifprod@font
    \scriptfont\itfam=\twelveit
      \scriptscriptfont\itfam=\tenit
  \else
    \scriptfont\itfam=\seventeenit
      \scriptscriptfont\itfam=\seventeenit
  \fi
  \textfont\bffam=\seventeenbf%
    \scriptfont\bffam=\twelvebf%
      \scriptscriptfont\bffam=\tenbf%
  \def\bf{\fam\bffam\seventeenbf}%
  \textfont\slfam=\seventeensl\def\sl{\fam\slfam\seventeensl}%
  \ifprod@font
    \scriptfont\slfam=\twelvesl
      \scriptscriptfont\slfam=\tensl
  \else
    \scriptfont\slfam=\seventeensl
      \scriptscriptfont\slfam=\seventeensl
  \fi
  \textfont\ttfam=\seventeentt\def\tt{\fam\ttfam\seventeentt}%
  \ifprod@font
    \scriptfont\ttfam=\twelvett
      \scriptscriptfont\ttfam=\tentt
  \else
    \scriptfont\ttfam=\seventeentt
      \scriptscriptfont\ttfam=\seventeentt
  \fi
  \textfont\scfam=\seventeencsc\def\sc{\fam\scfam\seventeencsc}%
  \ifprod@font
    \scriptfont\scfam=\twelvecsc
      \scriptscriptfont\scfam=\tencsc
  \else
    \scriptfont\scfam=\seventeencsc
      \scriptscriptfont\scfam=\seventeencsc
  \fi
  \textfont\sffam=\seventeensf\def\sf{\fam\sffam\seventeensf}%
  \ifprod@font
    \scriptfont\sffam=\twelvesf
      \scriptscriptfont\sffam=\tensf
  \else
    \scriptfont\sffam=\seventeensf
      \scriptscriptfont\sffam=\seventeensf
  \fi
  \textfont\mibfam=\seventeenmib
    \scriptfont\mibfam=\twelvemib
      \scriptscriptfont\mibfam=\tenmib
  \textfont\sybfam=\seventeensyb
    \scriptfont\sybfam=\twelvesyb
      \scriptscriptfont\sybfam=\tensyb
  \ifprod@font
    \textfont\xmfam=\seventeenxm
      \scriptfont\xmfam=\twelvexm
        \scriptscriptfont\xmfam=\tenxm
    \textfont\ymfam=\seventeenym
      \scriptfont\ymfam=\twelveym
        \scriptscriptfont\ymfam=\tenym
  \fi
  \def\oldstyle{\fam\@ne\seventeeni}%
  \def\boldstyle{\fam\mibfam\seventeenmib}%
  \b@ls{20pt}\rm%
}

\lineskip=1pt      \normallineskip=\lineskip
\lineskiplimit=\z@ \normallineskiplimit=\lineskiplimit



\def\la{\mathrel{\mathchoice {\vcenter{\offinterlineskip\halign{\hfil
$\displaystyle##$\hfil\cr<\cr\sim\cr}}}
{\vcenter{\offinterlineskip\halign{\hfil$\textstyle##$\hfil\cr
<\cr\sim\cr}}}
{\vcenter{\offinterlineskip\halign{\hfil$\scriptstyle##$\hfil\cr
<\cr\sim\cr}}}
{\vcenter{\offinterlineskip\halign{\hfil$\scriptscriptstyle##$\hfil\cr
<\cr\sim\cr}}}}}

\def\ga{\mathrel{\mathchoice {\vcenter{\offinterlineskip\halign{\hfil
$\displaystyle##$\hfil\cr>\cr\sim\cr}}}
{\vcenter{\offinterlineskip\halign{\hfil$\textstyle##$\hfil\cr
>\cr\sim\cr}}}
{\vcenter{\offinterlineskip\halign{\hfil$\scriptstyle##$\hfil\cr
>\cr\sim\cr}}}
{\vcenter{\offinterlineskip\halign{\hfil$\scriptscriptstyle##$\hfil\cr
>\cr\sim\cr}}}}}

\def\getsto{\mathrel{\mathchoice {\vcenter{\offinterlineskip
\halign{\hfil
$\displaystyle##$\hfil\cr\gets\cr\to\cr}}}
{\vcenter{\offinterlineskip\halign{\hfil$\textstyle##$\hfil\cr\gets
\cr\to\cr}}}
{\vcenter{\offinterlineskip\halign{\hfil$\scriptstyle##$\hfil\cr\gets
\cr\to\cr}}}
{\vcenter{\offinterlineskip\halign{\hfil$\scriptscriptstyle##$\hfil\cr
\gets\cr\to\cr}}}}}

\def\lid{\mathrel{\mathchoice {\vcenter{\offinterlineskip\halign{\hfil
$\displaystyle##$\hfil\cr<\cr\noalign{\vskip1.2pt}=\cr}}}
{\vcenter{\offinterlineskip\halign{\hfil$\textstyle##$\hfil\cr<\cr
\noalign{\vskip1.2pt}=\cr}}}
{\vcenter{\offinterlineskip\halign{\hfil$\scriptstyle##$\hfil\cr<\cr
\noalign{\vskip1pt}=\cr}}}
{\vcenter{\offinterlineskip\halign{\hfil$\scriptscriptstyle##$\hfil\cr
<\cr
\noalign{\vskip0.9pt}=\cr}}}}}

\def\gid{\mathrel{\mathchoice {\vcenter{\offinterlineskip\halign{\hfil
$\displaystyle##$\hfil\cr>\cr\noalign{\vskip1.2pt}=\cr}}}
{\vcenter{\offinterlineskip\halign{\hfil$\textstyle##$\hfil\cr>\cr
\noalign{\vskip1.2pt}=\cr}}}
{\vcenter{\offinterlineskip\halign{\hfil$\scriptstyle##$\hfil\cr>\cr
\noalign{\vskip1pt}=\cr}}}
{\vcenter{\offinterlineskip\halign{\hfil$\scriptscriptstyle##$\hfil\cr
>\cr
\noalign{\vskip0.9pt}=\cr}}}}}

\def\grole{\mathrel{\mathchoice {\vcenter{\offinterlineskip\halign{\hfil
$\displaystyle##$\hfil\cr>\cr\noalign{\vskip-1.5pt}<\cr}}}
{\vcenter{\offinterlineskip\halign{\hfil$\textstyle##$\hfil\cr
>\cr\noalign{\vskip-1.5pt}<\cr}}}
{\vcenter{\offinterlineskip\halign{\hfil$\scriptstyle##$\hfil\cr
>\cr\noalign{\vskip-1pt}<\cr}}}
{\vcenter{\offinterlineskip\halign{\hfil$\scriptscriptstyle##$\hfil\cr
>\cr\noalign{\vskip-0.5pt}<\cr}}}}}

\def\leogr{\mathrel{\mathchoice {\vcenter{\offinterlineskip\halign{\hfil
$\displaystyle##$\hfil\cr<\cr\noalign{\vskip-1.5pt}>\cr}}}
{\vcenter{\offinterlineskip\halign{\hfil$\textstyle##$\hfil\cr
<\cr\noalign{\vskip-1.5pt}>\cr}}}
{\vcenter{\offinterlineskip\halign{\hfil$\scriptstyle##$\hfil\cr
<\cr\noalign{\vskip-1pt}>\cr}}}
{\vcenter{\offinterlineskip\halign{\hfil$\scriptscriptstyle##$\hfil\cr
<\cr\noalign{\vskip-0.5pt}>\cr}}}}}

\def\loa{\mathrel{\mathchoice {\vcenter{\offinterlineskip\halign{\hfil
$\displaystyle##$\hfil\cr<\cr\approx\cr}}}
{\vcenter{\offinterlineskip\halign{\hfil$\textstyle##$\hfil\cr
<\cr\approx\cr}}}
{\vcenter{\offinterlineskip\halign{\hfil$\scriptstyle##$\hfil\cr
<\cr\approx\cr}}}
{\vcenter{\offinterlineskip\halign{\hfil$\scriptscriptstyle##$\hfil\cr
<\cr\approx\cr}}}}}

\def\goa{\mathrel{\mathchoice {\vcenter{\offinterlineskip\halign{\hfil
$\displaystyle##$\hfil\cr>\cr\approx\cr}}}
{\vcenter{\offinterlineskip\halign{\hfil$\textstyle##$\hfil\cr
>\cr\approx\cr}}}
{\vcenter{\offinterlineskip\halign{\hfil$\scriptstyle##$\hfil\cr
>\cr\approx\cr}}}
{\vcenter{\offinterlineskip\halign{\hfil$\scriptscriptstyle##$\hfil\cr
>\cr\approx\cr}}}}}

\def\diameter{{\ifmmode\mathchoice
{\ooalign{\hfil\hbox{$\displaystyle/$}\hfil\crcr
{\hbox{$\displaystyle\mathchar"20D$}}}}
{\ooalign{\hfil\hbox{$\textstyle/$}\hfil\crcr
{\hbox{$\textstyle\mathchar"20D$}}}}
{\ooalign{\hfil\hbox{$\scriptstyle/$}\hfil\crcr
{\hbox{$\scriptstyle\mathchar"20D$}}}}
{\ooalign{\hfil\hbox{$\scriptscriptstyle/$}\hfil\crcr
{\hbox{$\scriptscriptstyle\mathchar"20D$}}}}
\else{\ooalign{\hfil/\hfil\crcr\mathhexbox20D}}%
\fi}}

\def\sq{\ifmmode\squareforqed\else{\unskip\nobreak\hfil
\penalty50\hskip1em\null\nobreak\hfil\squareforqed
\parfillskip=0pt\finalhyphendemerits=0\endgraf}\fi}
\def\squareforqed{\hbox{\rlap{$\sqcap$}$\sqcup$}}


\def\bbbc{{\mathchoice {\setbox0=\hbox{$\displaystyle\rm C$}\hbox{\hbox
to0pt{\kern0.4\wd0\vrule height0.9\ht0\hss}\box0}}
{\setbox0=\hbox{$\textstyle\rm C$}\hbox{\hbox
to0pt{\kern0.4\wd0\vrule height0.9\ht0\hss}\box0}}
{\setbox0=\hbox{$\scriptstyle\rm C$}\hbox{\hbox
to0pt{\kern0.4\wd0\vrule height0.9\ht0\hss}\box0}}
{\setbox0=\hbox{$\scriptscriptstyle\rm C$}\hbox{\hbox
to0pt{\kern0.4\wd0\vrule height0.9\ht0\hss}\box0}}}}
\def\bbbq{{\mathchoice {\setbox0=\hbox{$\displaystyle\rm
Q$}\hbox{\raise
0.15\ht0\hbox to0pt{\kern0.4\wd0\vrule height0.8\ht0\hss}\box0}}
{\setbox0=\hbox{$\textstyle\rm Q$}\hbox{\raise
0.15\ht0\hbox to0pt{\kern0.4\wd0\vrule height0.8\ht0\hss}\box0}}
{\setbox0=\hbox{$\scriptstyle\rm Q$}\hbox{\raise
0.15\ht0\hbox to0pt{\kern0.4\wd0\vrule height0.7\ht0\hss}\box0}}
{\setbox0=\hbox{$\scriptscriptstyle\rm Q$}\hbox{\raise
0.15\ht0\hbox to0pt{\kern0.4\wd0\vrule height0.7\ht0\hss}\box0}}}}
\def\bbbt{{\mathchoice {\setbox0=\hbox{$\displaystyle\rm
T$}\hbox{\hbox to0pt{\kern0.3\wd0\vrule height0.9\ht0\hss}\box0}}
{\setbox0=\hbox{$\textstyle\rm T$}\hbox{\hbox
to0pt{\kern0.3\wd0\vrule height0.9\ht0\hss}\box0}}
{\setbox0=\hbox{$\scriptstyle\rm T$}\hbox{\hbox
to0pt{\kern0.3\wd0\vrule height0.9\ht0\hss}\box0}}
{\setbox0=\hbox{$\scriptscriptstyle\rm T$}\hbox{\hbox
to0pt{\kern0.3\wd0\vrule height0.9\ht0\hss}\box0}}}}
\def\bbbs{{\mathchoice
{\setbox0=\hbox{$\displaystyle     \rm S$}\hbox{\raise0.5\ht0\hbox
to0pt{\kern0.35\wd0\vrule height0.45\ht0\hss}\hbox
to0pt{\kern0.55\wd0\vrule height0.5\ht0\hss}\box0}}
{\setbox0=\hbox{$\textstyle        \rm S$}\hbox{\raise0.5\ht0\hbox
to0pt{\kern0.35\wd0\vrule height0.45\ht0\hss}\hbox
to0pt{\kern0.55\wd0\vrule height0.5\ht0\hss}\box0}}
{\setbox0=\hbox{$\scriptstyle      \rm S$}\hbox{\raise0.5\ht0\hbox
to0pt{\kern0.35\wd0\vrule height0.45\ht0\hss}\raise0.05\ht0\hbox
to0pt{\kern0.5\wd0\vrule height0.45\ht0\hss}\box0}}
{\setbox0=\hbox{$\scriptscriptstyle\rm S$}\hbox{\raise0.5\ht0\hbox
to0pt{\kern0.4\wd0\vrule height0.45\ht0\hss}\raise0.05\ht0\hbox
to0pt{\kern0.55\wd0\vrule height0.45\ht0\hss}\box0}}}}
\def\bbbz{{\mathchoice {\hbox{$\sf\textstyle Z\kern-0.4em Z$}}
{\hbox{$\sf\textstyle Z\kern-0.4em Z$}}
{\hbox{$\sf\scriptstyle Z\kern-0.3em Z$}}
{\hbox{$\sf\scriptscriptstyle Z\kern-0.2em Z$}}}}


\ifprod@font
  \mathchardef\la="3\@xm2E
  \mathchardef\getsto="3\@xm1C
  \mathchardef\lid="3\@xm35
  \mathchardef\grole="3\@xm3F
  \mathchardef\loa="3\@xm2F
  \mathchardef\ga="3\@xm26
  \mathchardef\gid="3\@xm3D
  \mathchardef\leogr="3\@xm37
  \mathchardef\goa="3\@xm27
  \mathchardef\sq="0\@xm03
%
%
\def\diameter{{%
  \ifmmode
    \mathchoice
    {\ooalign{\hfil\hbox{$\displaystyle/$}\hfil\crcr
    {\lower.2ex\hbox{$\displaystyle\mathchar"20D$}}}}%
    {\ooalign{\hfil\hbox{$\textstyle/$}\hfil\crcr
    {\lower.2ex\hbox{$\textstyle\mathchar"20D$}}}}%
    {\ooalign{\hfil\hbox{$\scriptstyle/$}\hfil\crcr
    {\lower.1ex\hbox{$\scriptstyle\mathchar"20D$}}}}%
    {\ooalign{\hfil\hbox{$\scriptscriptstyle/$}\hfil\crcr
    {\lower.1ex\hbox{$\scriptscriptstyle\mathchar"20D$}}}}%
  \else
    {\ooalign{\hfil/\hfil\crcr\lower.2ex\hbox{\mathhexbox20D}}}%
  \fi
}}
%
%

\def\bbbc{{\Bbb{C}}}
\def\bbbq{{\Bbb{Q}}}
\def\bbbt{{\Bbb{T}}}
\def\bbbs{{\Bbb{S}}}
\def\bbbz{{\Bbb{Z}}}
\fi


\ifprod@font
\mathchardef\boxdot="2\@xm00
\mathchardef\boxplus="2\@xm01
\mathchardef\boxtimes="2\@xm02
\mathchardef\square="0\@xm03
\mathchardef\blacksquare="0\@xm04
\mathchardef\centerdot="2\@xm05
\mathchardef\lozenge="0\@xm06
\mathchardef\blacklozenge="0\@xm07
\mathchardef\circlearrowright="3\@xm08
\mathchardef\circlearrowleft="3\@xm09
\mathchardef\rightleftharpoons="3\@xm0A
\mathchardef\leftrightharpoons="3\@xm0B
\mathchardef\boxminus="2\@xm0C
\mathchardef\Vdash="3\@xm0D
\mathchardef\Vvdash="3\@xm0E
\mathchardef\vDash="3\@xm0F
\mathchardef\twoheadrightarrow="3\@xm10
\mathchardef\twoheadleftarrow="3\@xm11
\mathchardef\leftleftarrows="3\@xm12
\mathchardef\rightrightarrows="3\@xm13
\mathchardef\upuparrows="3\@xm14
\mathchardef\downdownarrows="3\@xm15
\mathchardef\upharpoonright="3\@xm16

\mathchardef\downharpoonright="3\@xm17
\mathchardef\upharpoonleft="3\@xm18
\mathchardef\downharpoonleft="3\@xm19
\mathchardef\rightarrowtail="3\@xm1A
\mathchardef\leftarrowtail="3\@xm1B
\mathchardef\leftrightarrows="3\@xm1C
\mathchardef\rightleftarrows="3\@xm1D
\mathchardef\Lsh="3\@xm1E
\mathchardef\Rsh="3\@xm1F
\mathchardef\rightsquigarrow="3\@xm20
\mathchardef\leftrightsquigarrow="3\@xm21
\mathchardef\looparrowleft="3\@xm22
\mathchardef\looparrowright="3\@xm23
\mathchardef\circeq="3\@xm24
\mathchardef\succsim="3\@xm25
\mathchardef\gtrsim="3\@xm26
\mathchardef\gtrapprox="3\@xm27
\mathchardef\multimap="3\@xm28
\mathchardef\therefore="3\@xm29
\mathchardef\because="3\@xm2A
\mathchardef\doteqdot="3\@xm2B

\mathchardef\triangleq="3\@xm2C
\mathchardef\precsim="3\@xm2D
\mathchardef\lesssim="3\@xm2E
\mathchardef\lessapprox="3\@xm2F
\mathchardef\eqslantless="3\@xm30
\mathchardef\eqslantgtr="3\@xm31
\mathchardef\curlyeqprec="3\@xm32
\mathchardef\curlyeqsucc="3\@xm33
\mathchardef\preccurlyeq="3\@xm34
\mathchardef\leqq="3\@xm35
\mathchardef\leqslant="3\@xm36
\mathchardef\lessgtr="3\@xm37
\mathchardef\backprime="0\@xm38
\mathchardef\risingdotseq="3\@xm3A
\mathchardef\fallingdotseq="3\@xm3B
\mathchardef\succcurlyeq="3\@xm3C
\mathchardef\geqq="3\@xm3D
\mathchardef\geqslant="3\@xm3E
\mathchardef\gtrless="3\@xm3F
\mathchardef\sqsubset="3\@xm40
\mathchardef\sqsupset="3\@xm41
\mathchardef\vartriangleright="3\@xm42
\mathchardef\vartriangleleft="3\@xm43
\mathchardef\trianglerighteq="3\@xm44
\mathchardef\trianglelefteq="3\@xm45
\mathchardef\bigstar="0\@xm46
\mathchardef\between="3\@xm47
\mathchardef\blacktriangledown="0\@xm48
\mathchardef\blacktriangleright="3\@xm49
\mathchardef\blacktriangleleft="3\@xm4A
\mathchardef\vartriangle="0\@xm4D
\mathchardef\blacktriangle="0\@xm4E
\mathchardef\triangledown="0\@xm4F
\mathchardef\eqcirc="3\@xm50
\mathchardef\lesseqgtr="3\@xm51
\mathchardef\gtreqless="3\@xm52
\mathchardef\lesseqqgtr="3\@xm53
\mathchardef\gtreqqless="3\@xm54
\mathchardef\Rrightarrow="3\@xm56
\mathchardef\Lleftarrow="3\@xm57
\mathchardef\veebar="2\@xm59
\mathchardef\barwedge="2\@xm5A
\mathchardef\doublebarwedge="2\@xm5B
\mathchardef\angle="0\@xm5C
\mathchardef\measuredangle="0\@xm5D
\mathchardef\sphericalangle="0\@xm5E
\mathchardef\varpropto="3\@xm5F
\mathchardef\smallsmile="3\@xm60
\mathchardef\smallfrown="3\@xm61
\mathchardef\Subset="3\@xm62
\mathchardef\Supset="3\@xm63
\mathchardef\Cup="2\@xm64

\mathchardef\Cap="2\@xm65

\mathchardef\curlywedge="2\@xm66
\mathchardef\curlyvee="2\@xm67
\mathchardef\leftthreetimes="2\@xm68
\mathchardef\rightthreetimes="2\@xm69
\mathchardef\subseteqq="3\@xm6A
\mathchardef\supseteqq="3\@xm6B
\mathchardef\bumpeq="3\@xm6C
\mathchardef\Bumpeq="3\@xm6D
\mathchardef\lll="3\@xm6E

\mathchardef\ggg="3\@xm6F

\mathchardef\circledS="0\@xm73
\mathchardef\pitchfork="3\@xm74
\mathchardef\dotplus="2\@xm75
\mathchardef\backsim="3\@xm76
\mathchardef\backsimeq="3\@xm77
\mathchardef\complement="0\@xm7B
\mathchardef\intercal="2\@xm7C
\mathchardef\circledcirc="2\@xm7D
\mathchardef\circledast="2\@xm7E
\mathchardef\circleddash="2\@xm7F
\def\ulcorner{\delimiter"4\@xm70\@xm70 }
\def\urcorner{\delimiter"5\@xm71\@xm71 }
\def\llcorner{\delimiter"4\@xm78\@xm78 }
\def\lrcorner{\delimiter"5\@xm79\@xm79 }
\def\yen{\mathhexbox\@xm55 }
\def\checkmark{\mathhexbox\@xm58 }
\def\circledR{\mathhexbox\@xm72 }
\def\maltese{\mathhexbox\@xm7A }
\mathchardef\lvertneqq="3\@ym00
\mathchardef\gvertneqq="3\@ym01
\mathchardef\nleq="3\@ym02
\mathchardef\ngeq="3\@ym03
\mathchardef\nless="3\@ym04
\mathchardef\ngtr="3\@ym05
\mathchardef\nprec="3\@ym06
\mathchardef\nsucc="3\@ym07
\mathchardef\lneqq="3\@ym08
\mathchardef\gneqq="3\@ym09
\mathchardef\nleqslant="3\@ym0A
\mathchardef\ngeqslant="3\@ym0B
\mathchardef\lneq="3\@ym0C
\mathchardef\gneq="3\@ym0D
\mathchardef\npreceq="3\@ym0E
\mathchardef\nsucceq="3\@ym0F
\mathchardef\precnsim="3\@ym10
\mathchardef\succnsim="3\@ym11
\mathchardef\lnsim="3\@ym12
\mathchardef\gnsim="3\@ym13
\mathchardef\nleqq="3\@ym14
\mathchardef\ngeqq="3\@ym15
\mathchardef\precneqq="3\@ym16
\mathchardef\succneqq="3\@ym17
\mathchardef\precnapprox="3\@ym18
\mathchardef\succnapprox="3\@ym19
\mathchardef\lnapprox="3\@ym1A
\mathchardef\gnapprox="3\@ym1B
\mathchardef\nsim="3\@ym1C
\mathchardef\ncong="3\@ym1D

\mathchardef\varsubsetneq="3\@ym20
\mathchardef\varsupsetneq="3\@ym21
\mathchardef\nsubseteqq="3\@ym22
\mathchardef\nsupseteqq="3\@ym23
\mathchardef\subsetneqq="3\@ym24
\mathchardef\supsetneqq="3\@ym25
\mathchardef\varsubsetneqq="3\@ym26
\mathchardef\varsupsetneqq="3\@ym27
\mathchardef\subsetneq="3\@ym28
\mathchardef\supsetneq="3\@ym29
\mathchardef\nsubseteq="3\@ym2A
\mathchardef\nsupseteq="3\@ym2B
\mathchardef\nparallel="3\@ym2C
\mathchardef\nmid="3\@ym2D
\mathchardef\nshortmid="3\@ym2E
\mathchardef\nshortparallel="3\@ym2F
\mathchardef\nvdash="3\@ym30
\mathchardef\nVdash="3\@ym31
\mathchardef\nvDash="3\@ym32
\mathchardef\nVDash="3\@ym33
\mathchardef\ntrianglerighteq="3\@ym34
\mathchardef\ntrianglelefteq="3\@ym35
\mathchardef\ntriangleleft="3\@ym36
\mathchardef\ntriangleright="3\@ym37
\mathchardef\nleftarrow="3\@ym38
\mathchardef\nrightarrow="3\@ym39
\mathchardef\nLeftarrow="3\@ym3A
\mathchardef\nRightarrow="3\@ym3B
\mathchardef\nLeftrightarrow="3\@ym3C
\mathchardef\nleftrightarrow="3\@ym3D
\mathchardef\divideontimes="2\@ym3E
\mathchardef\varnothing="0\@ym3F
\mathchardef\nexists="0\@ym40
\mathchardef\mho="0\@ym66
\mathchardef\eth="0\@ym67
\mathchardef\eqsim="3\@ym68
\mathchardef\beth="0\@ym69
\mathchardef\gimel="0\@ym6A
\mathchardef\daleth="0\@ym6B
\mathchardef\lessdot="3\@ym6C
\mathchardef\gtrdot="3\@ym6D
\mathchardef\ltimes="2\@ym6E
\mathchardef\rtimes="2\@ym6F
\mathchardef\shortmid="3\@ym70
\mathchardef\shortparallel="3\@ym71
\mathchardef\smallsetminus="2\@ym72
\mathchardef\thicksim="3\@ym73
\mathchardef\thickapprox="3\@ym74
\mathchardef\approxeq="3\@ym75
\mathchardef\succapprox="3\@ym76
\mathchardef\precapprox="3\@ym77
\mathchardef\curvearrowleft="3\@ym78
\mathchardef\curvearrowright="3\@ym79
\mathchardef\digamma="0\@ym7A
\mathchardef\varkappa="0\@ym7B
\mathchardef\hslash="0\@ym7D
\mathchardef\hbar="0\@ym7E
\mathchardef\backepsilon="3\@ym7F


\def\Bbb{\ifmmode\let\next\Bbb@\else
\def\next{\errmessage{Use \string\Bbb\space only in math mode}}\fi\next}
\def\Bbb@#1{{\Bbb@@{#1}}}
\def\Bbb@@#1{\fam\ymfam#1}
\fi


\def\Nulle{0} 
\def\Afe{1}   
\def\Hae{2}   
\def\Hbe{3}   
\def\Hce{4}   
\def\Hde{5}   


\newcount\LastMac       \LastMac=\Nulle

\newskip\half      \half=5.5pt plus 1.5pt minus 2.25pt
\newskip\one       \one=11pt plus 3pt minus 5.5pt
\newskip\onehalf   \onehalf=16.5pt plus 5.5pt minus 8.25pt
\newskip\two       \two=22pt plus 5.5pt minus 11pt

\def\Half{\addvspace{\half}}
\def\One{\addvspace{\one}}
\def\OneHalf{\addvspace{\onehalf}}
\def\Two{\addvspace{\two}}


\def\Raggedright{
  \rightskip=\z@ plus \hsize\relax
}

\def\Fullout{
  \rightskip=\z@\relax
}

\def\Hang#1#2{
  \hangindent=#1%
  \hangafter=#2\relax
}


\newif\ifsp@page
\def\pagestyle#1{\csname ps@#1\endcsname}
\def\thispagestyle#1{\global\sp@pagetrue\gdef\sp@type{#1}}

\def\ps@titlepage{%
  \def\@oddhead{\eightpoint\noindent \the\CatchLine
    \ifprod@font\else\qquad Printed\ \today\fi \hfil}%
  \let\@evenhead=\@oddhead
}

\def\ps@headings{%
  \def\@oddhead{\elevenpoint\it\noindent
    \hfill\the\RightHeader\hskip1.5em\rm\folio}%
  \def\@evenhead{\elevenpoint\noindent
    \folio\hskip1.5em\it\the\LeftHeader\hfill}%
}

\def\ps@plate{%
  \def\@oddhead{\eightpoint\noindent\plt@cap\hfil}%
  \def\@evenhead{\eightpoint\noindent\plt@cap\hfil}%
}



\def\title#1{
  \bgroup
    \vbox to 8pt{\vss}%
    \seventeenpoint
    \Raggedright
    \noindent \strut{\bf #1}\par
  \egroup
}

\def\author#1{
  \bgroup
    \ifnum\LastMac=\Afe \OneHalf\else \vskip 21pt\fi
    \fourteenpoint
    \Raggedright
    \noindent \strut #1\par
    \vskip 3pt%
  \egroup
}

\def\affiliation#1{
  \bgroup
    \vskip -4pt%
    \eightpoint
    \Raggedright
    \noindent \strut {\it #1}\par
  \egroup
  \LastMac=\Afe\relax
}

\def\acceptedline#1{
  \bgroup
    \Two
    \eightpoint
    \Raggedright
    \noindent \strut #1\par
  \egroup
}

\long\def\abstract#1{%
  \bgroup
    \vskip 20pt%
    \everypar{\Hang{11pc}{0}}%
    \noindent{\ninebf ABSTRACT}\par
    \tenpoint
    \Fullout
    \noindent #1\par
  \egroup
}

\long\def\keywords#1{
  \bgroup
    \Half
    \everypar{\Hang{11pc}{0}}%
    \tenpoint
    \Fullout
    \noindent\hbox{\bf Key words:}\ #1\par
  \egroup
}


\def\maketitle{%
  \EndOpening
  \ifsinglecol \else \MakePage\fi
}


\def\pageoffset#1#2{\hoffset=#1\relax\voffset=#2\relax}


\def\Autonumber{
  \global\AutoNumbertrue  
}

\newif\ifAutoNumber \AutoNumberfalse
\newcount\Sec        
\newcount\SecSec
\newcount\SecSecSec

\Sec=\z@

\def\:{\let\@sptoken= } \:  
\def\:{\@xifnch} \expandafter\def\: {\futurelet\@tempc\@ifnch}

\def\@ifnextchar#1#2#3{%
  \let\@tempMACe #1%
  \def\@tempMACa{#2}%
  \def\@tempMACb{#3}%
  \futurelet \@tempMACc\@ifnch%
}

\def\@ifnch{%
\ifx \@tempMACc \@sptoken%
  \let\@tempMACd\@xifnch%
\else%
  \ifx \@tempMACc \@tempMACe%
    \let\@tempMACd\@tempMACa%
  \else%
    \let\@tempMACd\@tempMACb%
  \fi%
\fi%
\@tempMACd%
}

\def\@ifstar#1#2{\@ifnextchar *{\def\@tempMACa*{#1}\@tempMACa}{#2}}

\newskip\@tempskipb

\def\addvspace#1{%
  \ifvmode\else \endgraf\fi%
  \ifdim\lastskip=\z@%
    \vskip #1\relax%
  \else%
    \@tempskipb#1\relax\@xaddvskip%
  \fi%
}

\def\@xaddvskip{%
  \ifdim\lastskip<\@tempskipb%
    \vskip-\lastskip%
    \vskip\@tempskipb\relax%
  \else%
    \ifdim\@tempskipb<\z@%
      \ifdim\lastskip<\z@ \else%
        \advance\@tempskipb\lastskip%
        \vskip-\lastskip\vskip\@tempskipb%
      \fi%
    \fi%
  \fi%
}

\newskip\@tmpSKIP

\def\addpen#1{%
  \ifvmode
    \if@nobreak
    \else
      \ifdim\lastskip=\z@
        \penalty#1\relax
      \else
        \@tmpSKIP=\lastskip
        \vskip -\lastskip
        \penalty#1\vskip\@tmpSKIP
      \fi
    \fi
  \fi
}

\newcount\@clubpen   \@clubpen=\clubpenalty
\newif\if@nobreak    \@nobreakfalse

\def\@noafterindent{%
  \global\@nobreaktrue
  \everypar{\if@nobreak
              \global\@nobreakfalse
              \clubpenalty \@M
              {\setbox\z@\lastbox}%
              \LastMac=\Nulle\relax%
            \else
              \clubpenalty \@clubpen
              \everypar{}%
            \fi}
}

\newcount\gds@cbrk   \gds@cbrk=-300

\def\@nohdbrk{\interlinepenalty \@M\relax}

\let\@par=\par
\def\@restorepar{\def\par{\@par}}

\newif\if@endpe   \@endpefalse

\def\@doendpe{\@endpetrue \@nobreakfalse \LastMac=\Nulle\relax%
     \def\par{\@restorepar\everypar{}\par\@endpefalse}%
              \everypar{\setbox\z@\lastbox\everypar{}\@endpefalse}%
}

\def\section{\@ifstar{\@ssection}{\@section}}

\def\@section#1{
  \if@nobreak
    \everypar{}%
    \ifnum\LastMac=\Hae \addvspace{\half}\fi
  \else
    \addpen{\gds@cbrk}%
    \addvspace{\two}%
  \fi
  \bgroup
    \ninepoint\bf
    \Raggedright
    \ifAutoNumber
      \global\advance\Sec \@ne
      \noindent\@nohdbrk\number\Sec\hskip 1pc \uppercase{#1}\par
      \global\SecSec=\z@
    \else
      \noindent\@nohdbrk\uppercase{#1}\par
    \fi
  \egroup
  \nobreak
  \vskip\half
  \nobreak
  \@noafterindent
  \LastMac=\Hae\relax
}

\def\@ssection#1{
  \if@nobreak
    \everypar{}%
    \ifnum\LastMac=\Hae \addvspace{\half}\fi
  \else
    \addpen{\gds@cbrk}%
    \addvspace{\two}%
  \fi
  \bgroup
    \ninepoint\bf
    \Raggedright
    \noindent\@nohdbrk\uppercase{#1}\par
  \egroup
  \nobreak
  \vskip\half
  \nobreak
  \@noafterindent
  \LastMac=\Hae\relax
}

\def\subsection#1{
  \if@nobreak
    \everypar{}%
    \ifnum\LastMac=\Hae \addvspace{1pt plus 1pt minus .5pt}\fi
  \else
    \addpen{\gds@cbrk}%
    \addvspace{\onehalf}%
  \fi
  \bgroup
    \ninepoint\bf
    \Raggedright
    \ifAutoNumber
      \global\advance\SecSec \@ne
      \noindent\@nohdbrk\number\Sec.\number\SecSec \hskip 1pc\relax #1\par
      \global\SecSecSec=\z@
    \else
      \noindent\@nohdbrk #1\par
    \fi
  \egroup
  \nobreak
  \vskip\half
  \nobreak
  \@noafterindent
  \LastMac=\Hbe\relax
}

\def\subsubsection#1{
  \if@nobreak
    \everypar{}%
    \ifnum\LastMac=\Hbe \addvspace{1pt plus 1pt minus .5pt}\fi
  \else
    \addpen{\gds@cbrk}%
    \addvspace{\onehalf}%
  \fi
  \bgroup
    \ninepoint\it
    \Raggedright
    \ifAutoNumber
      \global\advance\SecSecSec \@ne
      \noindent\@nohdbrk\number\Sec.\number\SecSec.\number\SecSecSec
        \hskip 1pc\relax #1\par
    \else
      \noindent\@nohdbrk #1\par
    \fi
  \egroup
  \nobreak
  \vskip\half
  \nobreak
  \@noafterindent
  \LastMac=\Hce\relax
}

\def\paragraph#1{
  \if@nobreak
    \everypar{}%
  \else
    \addpen{\gds@cbrk}%
    \addvspace{\one}%
  \fi%
  \bgroup%
    \ninepoint\it
    \noindent #1\ \nobreak%
  \egroup
  \LastMac=\Hde\relax
  \ignorespaces
}




\def\beginlist{%
  \par\if@nobreak \else\addvspace{\half}\fi%
  \bgroup%
    \ninepoint
    \let\item=\list@item%
}

\def\list@item{%
  \par\noindent\hskip 1em\relax%
  \ignorespaces%
}

\def\endlist{\par\egroup\addvspace{\half}\@doendpe}


\def\beginrefs{%
  \par
  \bgroup
    \eightpoint
    \Raggedright
    \let\bibitem=\bib@item
}

\def\bib@item{%
  \par\parindent=1.5em\Hang{1.5em}{1}%
  \everypar={\Hang{1.5em}{1}\ignorespaces}%
  \noindent\ignorespaces
}

\def\endrefs{\par\egroup\@doendpe}


\newtoks\CatchLine

\def\@journal{Mon.\ Not.\ R.\ Astron.\ Soc.\ }  
\def\@pubyear{1994}        
\def\@pagerange{000--000}  
\def\@volume{000}          
\def\@microfiche{}         %

\def\pubyear#1{\gdef\@pubyear{#1}\@makecatchline}
\def\pagerange#1{\gdef\@pagerange{#1}\@makecatchline}
\def\volume#1{\gdef\@volume{#1}\@makecatchline}
\def\microfiche#1{\gdef\@microfiche{and Microfiche\ #1}\@makecatchline}

\def\@makecatchline{%
  \global\CatchLine{%
    {\rm \@journal {\bf \@volume},\ \@pagerange\ (\@pubyear)\ \@microfiche}}%
}

\@makecatchline 

\newtoks\LeftHeader
\def\shortauthor#1{
  \global\LeftHeader{#1}%
}

\newtoks\RightHeader
\def\shorttitle#1{
  \global\RightHeader{#1}%
}

\def\PageHead{
  \begingroup
    \ifsp@page
      \csname ps@\sp@type\endcsname
      \global\sp@pagefalse
    \fi
    \ifodd\pageno
      \let\the@head=\@oddhead
    \else
      \let\the@head=\@evenhead
    \fi
    \vbox to \z@{\vskip-22.5\p@%
      \hbox to \PageWidth{\vbox to8.5\p@{}%
        \the@head
      }%
    \vss}%
  \endgroup
  \nointerlineskip
}

\def\today{%
  \number\day\space
  \ifcase\month\or January\or February\or March\or April\or May\or June\or
    July\or August\or September\or October\or November\or December\fi
  \space\number\year%
}

\def\PageFoot{} 

\def\authorcomment#1{%
  \gdef\PageFoot{%
    \nointerlineskip%
    \vbox to 22pt{\vfil%
      \hbox to \PageWidth{\elevenpoint\noindent \hfil #1 \hfil}}%
  }%
}


\newif\ifplate@page
\newbox\plt@box

\def\beginplatepage{%
  \let\plate=\plate@head
  \let\caption=\fig@caption
  \global\setbox\plt@box=\vbox\bgroup
  \TEMPDIMEN=\PageWidth 
  \hsize=\PageWidth\relax
}

\def\endplatepage{\par\egroup\global\plate@pagetrue}
\def\plate@head#1{\gdef\plt@cap{#1}}


\def\letters{%
  \gdef\folio{\ifnum\pageno<\z@ L\romannumeral-\pageno
    \else L\number\pageno \fi}%
}


\everydisplay{\displaysetup}

\newif\ifeqno
\newif\ifleqno

\def\displaysetup#1$${%
 \displaytest#1\eqno\eqno\displaytest
}

\def\displaytest#1\eqno#2\eqno#3\displaytest{%
 \if!#3!\ldisplaytest#1\leqno\leqno\ldisplaytest
 \else\eqnotrue\leqnofalse\def\eqn{#2}\def\eq{#1}\fi
 \generaldisplay$$}

\def\ldisplaytest#1\leqno#2\leqno#3\ldisplaytest{%
 \def\eq{#1}%
 \if!#3!\eqnofalse\else\eqnotrue\leqnotrue
  \def\eqn{#2}\fi}

\def\generaldisplay{%
\ifeqno \ifleqno
   \hbox to \hsize{\noindent
     $\displaystyle\eq$\hfil$\displaystyle\eqn$}
  \else
    \hbox to \hsize{\noindent
     $\displaystyle\eq$\hfil$\displaystyle\eqn$}
  \fi
 \else
 \hbox to \hsize{\vbox{\noindent
  $\displaystyle\eq$\hfil}}
 \fi
}


\def\@notice{%
  \par\Two%
  \noindent{\b@ls{11pt}\ninerm This paper has been produced using the
    Blackwell Scientific Publications \TeX\ macros.\par}%
}

\outer\def\bye{\@notice\par\vfill\supereject\end}


\def\start@mess{%
  Monthly notices of the RAS journal style (\@typeface)\space
    v\@version,\space \@verdate.%
}

\everyjob{\Warn{\start@mess}}



\newif\if@debug \@debugfalse  

\def\Print#1{\if@debug\immediate\write16{#1}\else \fi}
\def\Warn#1{\immediate\write16{#1}}
\def\wlog#1{}

\newcount\Iteration 

\def\Single{0} \def\Double{1}                 
\def\Figure{0} \def\Table{1}                  

\def\InStack{0}  
\def\InZoneA{1}
\def\InZoneB{2}
\def\InZoneC{3}

\newcount\TEMPCOUNT 
\newdimen\TEMPDIMEN 
\newbox\TEMPBOX     
\newbox\VOIDBOX     

\newcount\LengthOfStack 
\newcount\MaxItems      
\newcount\StackPointer
\newcount\Point         
\newcount\NextFigure    
\newcount\NextTable     
\newcount\NextItem      

\newcount\StatusStack   
\newcount\NumStack      
\newcount\TypeStack     
\newcount\SpanStack     
\newcount\BoxStack      

\newcount\ItemSTATUS    
\newcount\ItemNUMBER    
\newcount\ItemTYPE      
\newcount\ItemSPAN      
\newbox\ItemBOX         
\newdimen\ItemSIZE      

\newdimen\PageHeight    
\newdimen\TextLeading   
\newdimen\Feathering    
\newcount\LinesPerPage  
\newdimen\ColumnWidth   
\newdimen\ColumnGap     
\newdimen\PageWidth     
\newdimen\BodgeHeight   
\newcount\Leading       

\newdimen\ZoneBSize  
\newdimen\TextSize   
\newbox\ZoneABOX     
\newbox\ZoneBBOX     
\newbox\ZoneCBOX     

\newif\ifFirstSingleItem
\newif\ifFirstZoneA
\newif\ifMakePageInComplete
\newif\ifMoreFigures \MoreFiguresfalse 
\newif\ifMoreTables  \MoreTablesfalse  

\newif\ifFigInZoneB 
\newif\ifFigInZoneC 
\newif\ifTabInZoneB 
\newif\ifTabInZoneC

\newif\ifZoneAFullPage

\newbox\MidBOX    
\newbox\LeftBOX
\newbox\RightBOX
\newbox\PageBOX   

\newif\ifLeftCOL  
\LeftCOLtrue

\newdimen\ZoneBAdjust

\newcount\ItemFits
\def\Yes{1}
\def\No{2}


\MaxItems=15
\NextFigure=\z@        
\NextTable=\@ne

\BodgeHeight=6pt
\TextLeading=11pt    
\Leading=11
\Feathering=\z@      
\LinesPerPage=61     
\topskip=\TextLeading
\ColumnWidth=20pc    
\ColumnGap=2pc       

\newskip\ItemSepamount  
\ItemSepamount=\TextLeading plus \TextLeading minus 4pt

\parskip=\z@ plus .1pt
\parindent=18pt
\widowpenalty=\z@
\clubpenalty=10000
\tolerance=1500
\hbadness=1500
\abovedisplayskip=6pt plus 2pt minus 2pt
\belowdisplayskip=6pt plus 2pt minus 2pt
\abovedisplayshortskip=6pt plus 2pt minus 2pt
\belowdisplayshortskip=6pt plus 2pt minus 2pt

\ninepoint 


\PageHeight=682pt

\PageWidth=2\ColumnWidth
\advance\PageWidth by \ColumnGap

\pagestyle{headings}




\newcount\DUMMY \StatusStack=\allocationnumber
\newcount\DUMMY \newcount\DUMMY \newcount\DUMMY
\newcount\DUMMY \newcount\DUMMY \newcount\DUMMY
\newcount\DUMMY \newcount\DUMMY \newcount\DUMMY
\newcount\DUMMY \newcount\DUMMY \newcount\DUMMY
\newcount\DUMMY \newcount\DUMMY \newcount\DUMMY

\newcount\DUMMY \NumStack=\allocationnumber
\newcount\DUMMY \newcount\DUMMY \newcount\DUMMY
\newcount\DUMMY \newcount\DUMMY \newcount\DUMMY
\newcount\DUMMY \newcount\DUMMY \newcount\DUMMY
\newcount\DUMMY \newcount\DUMMY \newcount\DUMMY
\newcount\DUMMY \newcount\DUMMY \newcount\DUMMY

\newcount\DUMMY \TypeStack=\allocationnumber
\newcount\DUMMY \newcount\DUMMY \newcount\DUMMY
\newcount\DUMMY \newcount\DUMMY \newcount\DUMMY
\newcount\DUMMY \newcount\DUMMY \newcount\DUMMY
\newcount\DUMMY \newcount\DUMMY \newcount\DUMMY
\newcount\DUMMY \newcount\DUMMY \newcount\DUMMY

\newcount\DUMMY \SpanStack=\allocationnumber
\newcount\DUMMY \newcount\DUMMY \newcount\DUMMY
\newcount\DUMMY \newcount\DUMMY \newcount\DUMMY
\newcount\DUMMY \newcount\DUMMY \newcount\DUMMY
\newcount\DUMMY \newcount\DUMMY \newcount\DUMMY
\newcount\DUMMY \newcount\DUMMY \newcount\DUMMY

\newbox\DUMMY   \BoxStack=\allocationnumber
\newbox\DUMMY   \newbox\DUMMY \newbox\DUMMY
\newbox\DUMMY   \newbox\DUMMY \newbox\DUMMY
\newbox\DUMMY   \newbox\DUMMY \newbox\DUMMY
\newbox\DUMMY   \newbox\DUMMY \newbox\DUMMY
\newbox\DUMMY   \newbox\DUMMY \newbox\DUMMY

\def\wlog{\immediate\write\m@ne}


\def\GetItemAll#1{%
 \GetItemSTATUS{#1}
 \GetItemNUMBER{#1}
 \GetItemTYPE{#1}
 \GetItemSPAN{#1}
 \GetItemBOX{#1}
}

\def\GetItemSTATUS#1{%
 \Point=\StatusStack
 \advance\Point by #1
 \global\ItemSTATUS=\count\Point
}

\def\GetItemNUMBER#1{%
 \Point=\NumStack
 \advance\Point by #1
 \global\ItemNUMBER=\count\Point
}

\def\GetItemTYPE#1{%
 \Point=\TypeStack
 \advance\Point by #1
 \global\ItemTYPE=\count\Point
}

\def\GetItemSPAN#1{%
 \Point\SpanStack
 \advance\Point by #1
 \global\ItemSPAN=\count\Point
}

\def\GetItemBOX#1{%
 \Point=\BoxStack
 \advance\Point by #1
 \global\setbox\ItemBOX=\vbox{\copy\Point}
 \global\ItemSIZE=\ht\ItemBOX
 \global\advance\ItemSIZE by \dp\ItemBOX
 \TEMPCOUNT=\ItemSIZE
 \divide\TEMPCOUNT by \Leading
 \divide\TEMPCOUNT by 65536
 \advance\TEMPCOUNT \@ne
 \ItemSIZE=\TEMPCOUNT pt
 \global\multiply\ItemSIZE by \Leading
}


\def\JoinStack{%
 \ifnum\LengthOfStack=\MaxItems 
  \Warn{WARNING: Stack is full...some items will be lost!}
 \else
  \Point=\StatusStack
  \advance\Point by \LengthOfStack
  \global\count\Point=\ItemSTATUS
  \Point=\NumStack
  \advance\Point by \LengthOfStack
  \global\count\Point=\ItemNUMBER
  \Point=\TypeStack
  \advance\Point by \LengthOfStack
  \global\count\Point=\ItemTYPE
  \Point\SpanStack
  \advance\Point by \LengthOfStack
  \global\count\Point=\ItemSPAN
  \Point=\BoxStack
  \advance\Point by \LengthOfStack
  \global\setbox\Point=\vbox{\copy\ItemBOX}
  \global\advance\LengthOfStack \@ne
  \ifnum\ItemTYPE=\Figure 
   \global\MoreFigurestrue
  \else
   \global\MoreTablestrue
  \fi
 \fi
}


\def\LeaveStack#1{%
 {\Iteration=#1
 \loop
 \ifnum\Iteration<\LengthOfStack
  \advance\Iteration \@ne
  \GetItemSTATUS{\Iteration}
   \advance\Point by \m@ne
   \global\count\Point=\ItemSTATUS
  \GetItemNUMBER{\Iteration}
   \advance\Point by \m@ne
   \global\count\Point=\ItemNUMBER
  \GetItemTYPE{\Iteration}
   \advance\Point by \m@ne
   \global\count\Point=\ItemTYPE
  \GetItemSPAN{\Iteration}
   \advance\Point by \m@ne
   \global\count\Point=\ItemSPAN
  \GetItemBOX{\Iteration}
   \advance\Point by \m@ne
   \global\setbox\Point=\vbox{\copy\ItemBOX}
 \repeat}
 \global\advance\LengthOfStack by \m@ne
}


\newif\ifStackNotClean

\def\CleanStack{%
 \StackNotCleantrue
 {\Iteration=\z@
  \loop
   \ifStackNotClean
    \GetItemSTATUS{\Iteration}
    \ifnum\ItemSTATUS=\InStack
     \advance\Iteration \@ne
     \else
      \LeaveStack{\Iteration}
    \fi
   \ifnum\LengthOfStack<\Iteration
    \StackNotCleanfalse
   \fi
 \repeat}
}


\def\FindItem#1#2{%
 \global\StackPointer=\m@ne 
 {\Iteration=\z@
  \loop
  \ifnum\Iteration<\LengthOfStack
   \GetItemSTATUS{\Iteration}
   \ifnum\ItemSTATUS=\InStack
    \GetItemTYPE{\Iteration}
    \ifnum\ItemTYPE=#1
     \GetItemNUMBER{\Iteration}
     \ifnum\ItemNUMBER=#2
      \global\StackPointer=\Iteration
      \Iteration=\LengthOfStack 
     \fi
    \fi
   \fi
  \advance\Iteration \@ne
 \repeat}
}


\def\FindNext{%
 \global\StackPointer=\m@ne 
 {\Iteration=\z@
  \loop
  \ifnum\Iteration<\LengthOfStack
   \GetItemSTATUS{\Iteration}
   \ifnum\ItemSTATUS=\InStack
    \GetItemTYPE{\Iteration}
   \ifnum\ItemTYPE=\Figure
    \ifMoreFigures
      \global\NextItem=\Figure
      \global\StackPointer=\Iteration
      \Iteration=\LengthOfStack 
    \fi
   \fi
   \ifnum\ItemTYPE=\Table
    \ifMoreTables
      \global\NextItem=\Table
      \global\StackPointer=\Iteration
      \Iteration=\LengthOfStack 
    \fi
   \fi
  \fi
  \advance\Iteration \@ne
 \repeat}
}


\def\ChangeStatus#1#2{%
 \Point=\StatusStack
 \advance\Point by #1
 \global\count\Point=#2
}



\def\Zone{\InZoneA}

\ZoneBAdjust=\z@

\def\MakePage{
 \global\ZoneBSize=\PageHeight
 \global\TextSize=\ZoneBSize
 \global\ZoneAFullPagefalse
 \global\topskip=\TextLeading
 \MakePageInCompletetrue
 \MoreFigurestrue
 \MoreTablestrue
 \FigInZoneBfalse
 \FigInZoneCfalse
 \TabInZoneBfalse
 \TabInZoneCfalse
 \global\FirstSingleItemtrue
 \global\FirstZoneAtrue
 \global\setbox\ZoneABOX=\box\VOIDBOX
 \global\setbox\ZoneBBOX=\box\VOIDBOX
 \global\setbox\ZoneCBOX=\box\VOIDBOX
 \loop
  \ifMakePageInComplete
 \FindNext
 \ifnum\StackPointer=\m@ne
  \NextItem=\m@ne
  \MoreFiguresfalse
  \MoreTablesfalse
 \fi
 \ifnum\NextItem=\Figure
   \FindItem{\Figure}{\NextFigure}
   \ifnum\StackPointer=\m@ne \global\MoreFiguresfalse
   \else
    \GetItemSPAN{\StackPointer}
    \ifnum\ItemSPAN=\Single \def\Zone{\InZoneB}\relax
     \ifFigInZoneC \global\MoreFiguresfalse\fi
    \else
     \def\Zone{\InZoneA}
     \ifFigInZoneB \def\Zone{\InZoneC}\fi
    \fi
   \fi
   \ifMoreFigures\Print{}\FigureItems\fi
 \fi
\ifnum\NextItem=\Table
   \FindItem{\Table}{\NextTable}
   \ifnum\StackPointer=\m@ne \global\MoreTablesfalse
   \else
    \GetItemSPAN{\StackPointer}
    \ifnum\ItemSPAN=\Single\relax
     \ifTabInZoneC \global\MoreTablesfalse\fi
    \else
     \def\Zone{\InZoneA}
     \ifTabInZoneB \def\Zone{\InZoneC}\fi
    \fi
   \fi
   \ifMoreTables\Print{}\TableItems\fi
 \fi
   \MakePageInCompletefalse 
   \ifMoreFigures\MakePageInCompletetrue\fi
   \ifMoreTables\MakePageInCompletetrue\fi
 \repeat
 \ifZoneAFullPage
  \global\TextSize=\z@
  \global\ZoneBSize=\z@
  \global\vsize=\z@\relax
  \global\topskip=\z@\relax
  \vbox to \z@{\vss}
  \eject
 \else
 \global\advance\ZoneBSize by -\ZoneBAdjust
 \global\vsize=\ZoneBSize
 \global\hsize=\ColumnWidth
 \global\ZoneBAdjust=\z@
 \ifdim\TextSize<23pt
 \Warn{}
 \Warn{* Making column fall short: TextSize=\the\TextSize *}
 \vskip-\lastskip\eject\fi
 \fi
}

\def\MakeRightCol{
 \global\TextSize=\ZoneBSize
 \MakePageInCompletetrue
 \MoreFigurestrue
 \MoreTablestrue
 \global\FirstSingleItemtrue
 \global\setbox\ZoneBBOX=\box\VOIDBOX
 \def\Zone{\InZoneB}
 \loop
  \ifMakePageInComplete
 \FindNext
 \ifnum\StackPointer=\m@ne
  \NextItem=\m@ne
  \MoreFiguresfalse
  \MoreTablesfalse
 \fi
 \ifnum\NextItem=\Figure
   \FindItem{\Figure}{\NextFigure}
   \ifnum\StackPointer=\m@ne \MoreFiguresfalse
   \else
    \GetItemSPAN{\StackPointer}
    \ifnum\ItemSPAN=\Double\relax
     \MoreFiguresfalse\fi
   \fi
   \ifMoreFigures\Print{}\FigureItems\fi
 \fi
 \ifnum\NextItem=\Table
   \FindItem{\Table}{\NextTable}
   \ifnum\StackPointer=\m@ne \MoreTablesfalse
   \else
    \GetItemSPAN{\StackPointer}
    \ifnum\ItemSPAN=\Double\relax
     \MoreTablesfalse\fi
   \fi
   \ifMoreTables\Print{}\TableItems\fi
 \fi
   \MakePageInCompletefalse 
   \ifMoreFigures\MakePageInCompletetrue\fi
   \ifMoreTables\MakePageInCompletetrue\fi
 \repeat
 \ifZoneAFullPage
  \global\TextSize=\z@
  \global\ZoneBSize=\z@
  \global\vsize=\z@\relax
  \global\topskip=\z@\relax
  \vbox to \z@{\vss}
  \eject
 \else
 \global\vsize=\ZoneBSize
 \global\hsize=\ColumnWidth
 \ifdim\TextSize<23pt
 \Warn{}
 \Warn{* Making column fall short: TextSize=\the\TextSize *}
 \vskip-\lastskip\eject\fi
\fi
}

\def\FigureItems{
 \Print{Considering...}
 \ShowItem{\StackPointer}
 \GetItemBOX{\StackPointer} 
 \GetItemSPAN{\StackPointer}
  \CheckFitInZone 
  \ifnum\ItemFits=\Yes
   \ifnum\ItemSPAN=\Single
     \ChangeStatus{\StackPointer}{\InZoneB} 
     \global\FigInZoneBtrue
     \ifFirstSingleItem
      \hbox{}\vskip-\BodgeHeight
     \global\advance\ItemSIZE by \TextLeading
     \fi
     \unvbox\ItemBOX\ItemSep
     \global\FirstSingleItemfalse
     \global\advance\TextSize by -\ItemSIZE
     \global\advance\TextSize by -\TextLeading
   \else
    \ifFirstZoneA
     \global\advance\ItemSIZE by \TextLeading
     \global\FirstZoneAfalse\fi
    \global\advance\TextSize by -\ItemSIZE
    \global\advance\TextSize by -\TextLeading
    \global\advance\ZoneBSize by -\ItemSIZE
    \global\advance\ZoneBSize by -\TextLeading
    \ifFigInZoneB\relax
     \else
     \ifdim\TextSize<3\TextLeading
     \global\ZoneAFullPagetrue
     \fi
    \fi
    \ChangeStatus{\StackPointer}{\Zone}
    \ifnum\Zone=\InZoneC \global\FigInZoneCtrue\fi
  \fi
   \Print{TextSize=\the\TextSize}
   \Print{ZoneBSize=\the\ZoneBSize}
  \global\advance\NextFigure \@ne
   \Print{This figure has been placed.}
  \else
   \Print{No space available for this figure...holding over.}
   \Print{}
   \global\MoreFiguresfalse
  \fi
}

\def\TableItems{
 \Print{Considering...}
 \ShowItem{\StackPointer}
 \GetItemBOX{\StackPointer} 
 \GetItemSPAN{\StackPointer}
  \CheckFitInZone 
  \ifnum\ItemFits=\Yes
   \ifnum\ItemSPAN=\Single
    \ChangeStatus{\StackPointer}{\InZoneB}
     \global\TabInZoneBtrue
     \ifFirstSingleItem
      \hbox{}\vskip-\BodgeHeight
     \global\advance\ItemSIZE by \TextLeading
     \fi
     \unvbox\ItemBOX\ItemSep
     \global\FirstSingleItemfalse
     \global\advance\TextSize by -\ItemSIZE
     \global\advance\TextSize by -\TextLeading
   \else
    \ifFirstZoneA
    \global\advance\ItemSIZE by \TextLeading
    \global\FirstZoneAfalse\fi
    \global\advance\TextSize by -\ItemSIZE
    \global\advance\TextSize by -\TextLeading
    \global\advance\ZoneBSize by -\ItemSIZE
    \global\advance\ZoneBSize by -\TextLeading
    \ifFigInZoneB\relax
     \else
     \ifdim\TextSize<3\TextLeading
     \global\ZoneAFullPagetrue
     \fi
    \fi
    \ChangeStatus{\StackPointer}{\Zone}
    \ifnum\Zone=\InZoneC \global\TabInZoneCtrue\fi
   \fi
  \global\advance\NextTable \@ne
   \Print{This table has been placed.}
  \else
  \Print{No space available for this table...holding over.}
   \Print{}
   \global\MoreTablesfalse
  \fi
}


\def\CheckFitInZone{%
{\advance\TextSize by -\ItemSIZE
 \advance\TextSize by -\TextLeading
 \ifFirstSingleItem
  \advance\TextSize by \TextLeading
 \fi
 \ifnum\Zone=\InZoneA\relax
  \else \advance\TextSize by -\ZoneBAdjust
 \fi
 \ifdim\TextSize<3\TextLeading \global\ItemFits=\No
 \else \global\ItemFits=\Yes\fi}
}

\def\BeginOpening{%
  \thispagestyle{titlepage}%
  \global\setbox\ItemBOX=\vbox\bgroup%
    \hsize=\PageWidth%
    \hrule height \z@
    \ifsinglecol\vskip 6pt\fi 
}

\let\begintopmatter=\BeginOpening  

\def\EndOpening{%
  \One
  \egroup
  \ifsinglecol
    \box\ItemBOX%
    \vskip\TextLeading plus 2\TextLeading
    \@noafterindent
  \else
    \ItemNUMBER=\z@%
    \ItemTYPE=\Figure
    \ItemSPAN=\Double
    \ItemSTATUS=\InStack
    \JoinStack
  \fi
}


\newif\if@here  \@herefalse

\def\no@float{\global\@heretrue}
\let\nofloat=\relax 

\def\beginfigure{%
  \@ifstar{\global\@dfloattrue \@bfigure}{\global\@dfloatfalse \@bfigure}%
}

\def\@bfigure#1{%
  \par
  \if@dfloat
    \ItemSPAN=\Double
    \TEMPDIMEN=\PageWidth
  \else
    \ItemSPAN=\Single
    \TEMPDIMEN=\ColumnWidth
  \fi
  \ifsinglecol
    \TEMPDIMEN=\PageWidth
  \else
    \ItemSTATUS=\InStack
    \ItemNUMBER=#1%
    \ItemTYPE=\Figure
  \fi
  \bgroup
    \hsize=\TEMPDIMEN
    \global\setbox\ItemBOX=\vbox\bgroup
      \eightpoint\nostb@ls{10pt}%
      \let\caption=\fig@caption
      \ifsinglecol \let\nofloat=\no@float\fi
}

\def\fig@caption#1{%
  \vskip 5.5pt plus 6pt%
  \bgroup 
    \eightpoint\nostb@ls{10pt}%
    \setbox\TEMPBOX=\hbox{#1}%
    \ifdim\wd\TEMPBOX>\TEMPDIMEN
      \noindent \unhbox\TEMPBOX\par
    \else
      \hbox to \hsize{\hfil\unhbox\TEMPBOX\hfil}%
    \fi
  \egroup
}

\def\endfigure{%
  \par\egroup 
  \egroup
  \ifsinglecol
    \if@here \midinsert\global\@herefalse\else \topinsert\fi
      \unvbox\ItemBOX
    \endinsert
  \else
    \JoinStack
    \Print{Processing source for figure \the\ItemNUMBER}%
  \fi
}


\newbox\tab@cap@box
\def\tab@caption#1{\global\setbox\tab@cap@box=\hbox{#1\par}}

\newtoks\tab@txt@toks
\long\def\tab@txt#1{\global\tab@txt@toks={#1}\global\table@txttrue}

\newif\iftable@txt  \table@txtfalse
\newif\if@dfloat    \@dfloatfalse

\def\begintable{%
  \@ifstar{\global\@dfloattrue \@btable}{\global\@dfloatfalse \@btable}%
}

\def\@btable#1{%
  \par
  \if@dfloat
    \ItemSPAN=\Double
    \TEMPDIMEN=\PageWidth
  \else
    \ItemSPAN=\Single
    \TEMPDIMEN=\ColumnWidth
  \fi
  \ifsinglecol
    \TEMPDIMEN=\PageWidth
  \else
    \ItemSTATUS=\InStack
    \ItemNUMBER=#1%
    \ItemTYPE=\Table
  \fi
  \bgroup
    \eightpoint\nostb@ls{10pt}%
    \global\setbox\ItemBOX=\vbox\bgroup
      \let\caption=\tab@caption
      \let\tabletext=\tab@txt
      \ifsinglecol \let\nofloat=\no@float\fi
}

\def\endtable{%
  \par\egroup 
  \egroup
  \setbox\TEMPBOX=\hbox to \TEMPDIMEN{%
    \hss
    \vbox{%
      \hsize=\wd\ItemBOX
      \ifvoid\tab@cap@box
      \else
        \noindent\unhbox\tab@cap@box
        \vskip 5.5pt plus 6pt%
      \fi
      \box\ItemBOX
      \iftable@txt
        \vskip 10pt%
        \eightpoint\nostb@ls{10pt}%
        \noindent\the\tab@txt@toks
        \global\table@txtfalse
      \fi
    }%
    \hss
  }%
  \ifsinglecol
    \if@here \midinsert\global\@herefalse\else \topinsert\fi
      \box\TEMPBOX
    \endinsert
  \else
    \global\setbox\ItemBOX=\box\TEMPBOX
    \JoinStack
    \Print{Processing source for table \the\ItemNUMBER}%
  \fi
}

\def\UnloadZoneA{%
\FirstZoneAtrue
 \Iteration=\z@
  \loop
   \ifnum\Iteration<\LengthOfStack
    \GetItemSTATUS{\Iteration}
    \ifnum\ItemSTATUS=\InZoneA
     \GetItemBOX{\Iteration}
     \ifFirstZoneA \vbox to \BodgeHeight{\vfil}%
     \FirstZoneAfalse\fi
     \unvbox\ItemBOX\ItemSep
     \LeaveStack{\Iteration}
     \else
     \advance\Iteration \@ne
   \fi
 \repeat
}

\def\UnloadZoneC{%
\Iteration=\z@
  \loop
   \ifnum\Iteration<\LengthOfStack
    \GetItemSTATUS{\Iteration}
    \ifnum\ItemSTATUS=\InZoneC
     \GetItemBOX{\Iteration}
     \ItemSep\unvbox\ItemBOX
     \LeaveStack{\Iteration}
     \else
     \advance\Iteration \@ne
   \fi
 \repeat
}


\def\ShowItem#1{
  {\GetItemAll{#1}
  \Print{\the#1:
  {TYPE=\ifnum\ItemTYPE=\Figure Figure\else Table\fi}
  {NUMBER=\the\ItemNUMBER}
  {SPAN=\ifnum\ItemSPAN=\Single Single\else Double\fi}
  {SIZE=\the\ItemSIZE}}}
}

\def\ShowStack{%
 \Print{}
 \Print{LengthOfStack = \the\LengthOfStack}
 \ifnum\LengthOfStack=\z@ \Print{Stack is empty}\fi
 \Iteration=\z@
 \loop
 \ifnum\Iteration<\LengthOfStack
  \ShowItem{\Iteration}
  \advance\Iteration \@ne
 \repeat
}

\def\B#1#2{%
\hbox{\vrule\kern-0.4pt\vbox to #2{%
\hrule width #1\vfill\hrule}\kern-0.4pt\vrule}
}


\newif\ifsinglecol   \singlecolfalse

\def\onecolumn{%
  \global\output={\singlecoloutput}%
  \global\hsize=\PageWidth
  \global\vsize=\PageHeight
  \global\ColumnWidth=\hsize
  \global\TextLeading=12pt
  \global\Leading=12
  \global\singlecoltrue
  \global\let\onecolumn=\relax
  \global\let\footnote=\sing@footnote
  \global\let\vfootnote=\sing@vfootnote
  \ninepoint 
  \message{(Single column)}%
}

\def\singlecoloutput{%
  \shipout\vbox{\PageHead\pagebody\PageFoot}%
  \advancepageno
  \ifplate@page
    \shipout\vbox{%
      \sp@pagetrue
      \def\sp@type{plate}%
      \global\plate@pagefalse
      \PageHead\vbox to \PageHeight{\unvbox\plt@box\vfil}\PageFoot%
    }%
    \message{[plate]}%
    \advancepageno
  \fi
  \ifnum\outputpenalty>-\@MM \else\dosupereject\fi%
}

\def\ItemSep{\vskip\ItemSepamount\relax}

\def\ItemSepbreak{\par\ifdim\lastskip<\ItemSepamount
  \removelastskip\penalty-200\ItemSep\fi%
}


\let\@@endinsert=\endinsert 

\def\endinsert{\egroup 
  \if@mid \dimen@\ht\z@ \advance\dimen@\dp\z@ \advance\dimen@12\p@
    \advance\dimen@\pagetotal \advance\dimen@-\pageshrink
    \ifdim\dimen@>\pagegoal\@midfalse\p@gefalse\fi\fi
  \if@mid \ItemSep\box\z@\ItemSepbreak
  \else\insert\topins{\penalty100 
    \splittopskip\z@skip
    \splitmaxdepth\maxdimen \floatingpenalty\z@
    \ifp@ge \dimen@\dp\z@
    \vbox to\vsize{\unvbox\z@\kern-\dimen@}
    \else \box\z@\nobreak\ItemSep\fi}\fi\endgroup%
}


\def\gobbleone#1{}
\def\gobbletwo#1#2{}
\let\footnote=\gobbletwo 
\let\vfootnote=\gobbleone

\def\sing@footnote#1{\let\@sf\empty 
  \ifhmode\edef\@sf{\spacefactor\the\spacefactor}\/\fi
  \hbox{$^{\hbox{\eightpoint #1}}$}\@sf\sing@vfootnote{#1}%
}

\def\sing@vfootnote#1{\insert\footins\bgroup\eightpoint\b@ls{9pt}%
  \interlinepenalty\interfootnotelinepenalty
  \splittopskip\ht\strutbox 
  \splitmaxdepth\dp\strutbox \floatingpenalty\@MM
  \leftskip\z@skip \rightskip\z@skip \spaceskip\z@skip \xspaceskip\z@skip
  \noindent $^{\scriptstyle\hbox{#1}}$\hskip 4pt%
    \footstrut\futurelet\next\fo@t%
}

\def\footnoterule{\kern-3\p@ \hrule height \z@ \kern 3\p@}

\skip\footins=19.5pt plus 12pt minus 1pt
\count\footins=1000
\dimen\footins=\maxdimen


\def\landscape{%
  \global\TEMPDIMEN=\PageWidth
  \global\PageWidth=\PageHeight
  \global\PageHeight=\TEMPDIMEN
  \global\let\landscape=\relax
  \onecolumn
  \message{(landscape)}%
  \raggedbottom
}


\output{%
  \ifLeftCOL
    \global\setbox\LeftBOX=\vbox to \ZoneBSize{\box255\unvbox\ZoneBBOX}%
    \global\LeftCOLfalse
    \MakeRightCol
  \else
    \setbox\RightBOX=\vbox to \ZoneBSize{\box255\unvbox\ZoneBBOX}%
    \setbox\MidBOX=\hbox{\box\LeftBOX\hskip\ColumnGap\box\RightBOX}%
    \setbox\PageBOX=\vbox to \PageHeight{%
      \UnloadZoneA\box\MidBOX\UnloadZoneC}%
    \shipout\vbox{\PageHead\box\PageBOX\PageFoot}%
    \advancepageno
    \ifplate@page
      \shipout\vbox{%
        \sp@pagetrue
        \def\sp@type{plate}%
        \global\plate@pagefalse
        \PageHead\vbox to \PageHeight{\unvbox\plt@box\vfil}\PageFoot%
      }%
      \message{[plate]}%
      \advancepageno
    \fi
    \global\LeftCOLtrue
    \CleanStack
    \MakePage
  \fi
}


\Warn{\start@mess}

\def\mnmacrosloaded{} 

\catcode `\@=12 



%% file: fig000.tex
\newcount\fignumber\fignumber=0\relax
\def\fign{\advance\fignumber by 1}

\def\figmod#1
{
	\ifnum#1>0
		\input figyes
	\else
		\input fignot
	\fi
}

\def\em{\ifdim\fontdimen1\font>\z@ \rm\else\it\fi}

%% file: figyes.tex
\input psfig

\def\putfig#1#2#3{\setbox20=\hbox
	{
	\psfig{file=#1,height=#2cm,clip=,angle=#3}
	}
	\centerline{$\vcenter{\box20}$}}

\def\putfigl#1#2#3{\setbox20=\hbox
	{
	\psfig{file=#1,height=#2cm,angle=#3}
	}
	\centerline{$\vcenter{\box20}$}}